\documentclass[10pt]{article}
\usepackage{graphicx}
\usepackage{amsmath}
\usepackage{amssymb}
\usepackage{caption2}
\usepackage{xcolor}
\begin{document}
\bibliographystyle{prsty}
\begin{center}
{\large {\bf \sc{  Analysis of  the hidden-charm pentaquark candidates in the $J/\psi \Lambda$ mass spectrum  via the  QCD sum rules }}} \\[2mm]
Zhi-Gang Wang \footnote{E-mail: zgwang@aliyun.com.  }, Qi Xin     \\
 Department of Physics, North China Electric Power University, Baoding 071003, P. R. China
\end{center}

\begin{abstract}
In this work, we distinguish the isospin  for the first time and  study the diquark-diquark-antiquark type $udsc\bar{c}$ pentaquark states with zero isospin via  the QCD sum rules systematically. We distinguish  contributions of the pentaquark states with negative parity from   positive parity unambiguously and obtain clean QCD sum rules for the  pentaquark states with negative parity.  Then we adopt  the modified energy scale formula to choose  the optimal  energy scales of the QCD spectral densities, and  obtain the mass spectrum of the $udsc\bar{c}$ pentaquark states with the quantum numbers $I=0$ and $J^{P}={\frac{1}{2}}^-$, ${\frac{3}{2}}^-$, ${\frac{5}{2}}^-$, which could interpret the $P_{cs}(4338)$ and $P_{cs}(4459)$ in the $J/\psi \Lambda$ mass spectrum naturally.
\end{abstract}



\section{Introduction}
In 2020, the LHCb collaboration reported an evidence of a hidden-charm pentaquark candidate $P_{cs}(4459)$ with the strangeness $S=-1$ in the $J/\psi \Lambda$ mass spectrum with  a statistical significance of  $3.1\sigma$ in the $\Xi_b^- \to J/\psi K^- \Lambda$ decays  \cite{LHCb-Pcs4459-2012},
the Breit-Wigner mass and width are
\begin{flalign}
 &P_{cs}(4459) : M = 4458.8 \pm 2.9 {}^{+4.7}_{-1.1} \mbox{ MeV}\, , \, \Gamma = 17.3 \pm 6.5 {}^{+8.0}_{-5.7} \mbox{ MeV} \, ,
\end{flalign}
but the spin and parity have not been determined yet.

In 2022, the LHCb collaboration observed an evidence for a new structure $P_{cs}(4338)$ in the $J/\psi \Lambda$ mass distribution in the $B^- \to J/\psi \Lambda \bar{p}$ decays \cite{LHCb-Pcs4338}. The measured  Breit-Wigner mass and width are $4338.2\pm0.7\pm0.4\,\rm{MeV}$ and $7.0\pm1.2\pm1.3\,\rm{MeV}$ respectively and the favored  spin-parity is $J^P={\frac{1}{2}}^-$.

Recently, the Belle and Belle-II collaborations observed the $\Upsilon(1{\rm S }, 2{\rm S})$ inclusive decays to the final states $J/\psi\Lambda$, and found  an evidence of the $P_{cs}(4459)$ state with a local significance of $3.3\,\sigma$, the measured mass and width
are $(4471.7 \pm 4.8 \pm 0.6)~\rm{MeV}$ and $(22 \pm 13 \pm 3)~\rm{MeV}$,
respectively \cite{Belle-Pcs4338-Pcs4459}.

The $P_{cs}(4338)$ and $P_{cs}(4459)$ are observed in the $J/\psi \Lambda$ invariant mass distribution,  they have the isospin $I=0$, as the strong decays conserve the isospin in most cases, the observation of their isospin cousins are of crucial importance.
The  possible  assignments are diquark-diquark-antiquark type pentaquark states \cite{WangZG-Pcs4459-333,Pcs4459-333-SR-Azizi,WangZG-Review,Pcs4459-mole-333-SR},  molecular   states \cite{Pcs4459-mole-333-SR,Pcs4459-mole-HXChen-SR,Pcs4459-mole-WangZG-SR,
Pcs4338-mole-XWWang,Pcs4459-mole-MJYan,
Pcs4459-mole-JHe,Pcs4459-mole-BWang,Pcs4459-mole-RChen,Pcs4459-mole-RChen-2,
Pcs4459-mole-LSGeng,Pcs4459-mole-CWXiao,Pcs4459-mole-DYChen,Pcs4459-mole-ZWLiu,
  Pcs4459-mole-FLWang,Pcs4459-mole-MLDu,Pcs4459-mole-Feijoo,Pcs4459-mole-JXLu,
  Pcs4459-mole-Nakamura,Pcs4459-mole-Ozdem-SR,Pcs4459-mole-XHu,Pcs4459-mole-FYang,
  Pcs4338-mole-LMeng,Pcs4338-mole-Ortega,Pcs4338-mole-JTZhu,Pcs4338-mole-FKGuo},  etc.

In 2021, the LHCb collaboration observed evidences  for a new structure in the $J/\psi p$ and $J/\psi \bar{p}$ systems with a mass of $4337 \ ^{+7}_{-4} \ ^{+2}_{-2}~\text{MeV}$ and a width of $29 \ ^{+26}_{-12} \ ^{+14}_{-14}~\text{MeV}$  with a significance in the range of 3.1 to 3.7$\sigma$,  which depend on the assigned $J^P$ hypothesis \cite{LHCb-Pc4337}. Although  it lies not far way from  the  $\bar{D}^* \Lambda_c$,
  $\bar{D} \Sigma_c$ and $\bar{D} \Sigma_c^*$ thresholds, it does not lie just in any baryon-meson  threshold, it is difficult to assign it as a molecular state without introducing large coupled channel effects. The molecule scenario still needs fine-tuning, we expect to obtain a suitable and uniform scheme to accommodate  all the existing pentaquark candidates.

The QCD sum rules approach is a powerful theoretical tool in exploring the exotic states, such as the tetraquark states, pentaquark states, molecular states, etc \cite{WangZG-Review,Nielsen-review}.
 In Refs.\cite{Pcs4338-mole-XWWang,WangXW-Pc-mole-SCPMA,WangXW-Pc-mole-IJMPA}, we
 distinguish the isospin for the first time, and study the color singlet-singlet type pentaquark states without strangeness and with strangeness in the framework of the QCD sum rules in a comprehensive way in our unique scheme, and observe that the observed pentaquark candidates  except for the $P_c(4337)$ could find their suitable positions in the scenario of molecules, for example,  the $P_{cs}(4459)$ can be assigned  as the $\bar{D}\Xi_c^{*}$ or $\bar{D}^*\Xi_c$ molecular state with the quantum numbers $(I,J^P)=(0,{\frac{3}{2}}^-)$, the $P_{cs}(4338)$  can be assigned as  the $\bar{D}\Xi_c$   molecular state with the quantum numbers  $(I,J^P)=(0,{\frac{1}{2}}^-)$, the observation  of their isospin cousins would shed light on the nature of those  pentaquark candidates.

 In Refs.\cite{Wang1508-EPJC,WangHuang-EPJC-1508-12,WangZG-EPJC-1509-12,
 WangZG-NPB-1512-32,WangZhang-APPB}, we adopt the pentaquark scenario and study the diquark-diquark-antiquark type hidden-charm pentaquark states with the spin-parity  $J^P={\frac{1}{2}}^\pm$, ${\frac{3}{2}}^\pm$, ${\frac{5}{2}}^\pm$  and the strangeness   $S=0,\,-1,\,-2,\,-3$ in the framework of  the QCD sum rules  systematically. Considering the tedious calculations in performing the operator product expansion, we only calculate the vacuum condensates up to  dimension 10, and the Borel platforms are not flat enough.

After the discovery of the $P_c(4312)$, the lowest pentaquark candidate with the valence quarks $uudc\bar{c}$,   we updated the old analysis and calculated  the   vacuum condensates up to dimension $13$  consistently,  and  restudied the ground state mass spectrum of the diquark-diquark-antiquark type $uudc\bar{c}$ pentaquark states, assigned the $P_c(4312)$, $P_c(4380)$, $P_c(4440)$ and $P_c(4457)$ in a reasonable way \cite{WZG-penta-IJMPA}.
 More importantly, we predicted a $uudc\bar{c}$ pentaquark state with the quantum numbers $(I,J^P)=(\frac{3}{2},{\frac{1}{2}}^-)$ and mass $4.34\pm0.14\,\rm{GeV}$, the corresponding $uudc\bar{c}$ pentaquark state with the quantum numbers $(I,J^P)=(\frac{1}{2},{\frac{1}{2}}^-)$ would like have slightly smaller mass and account for the $P_c(4337)$ reasonably \cite{WangZG-Review}.

 After the discovery of the $P_{cs}(4459)$, we studied the possibility of assigning it as the isospin cousin of the $P_{c}(4312)$ by taking account of the light-flavor $SU(3)$ breaking effects \cite{WangZG-Pcs4459-333}. Now we extend our previous works to study the diquark-diquark-antiquark type $udsc\bar{c}$ with the isospin $I=0$ and spin-parity $J^P={\frac{1}{2}}^-$, ${\frac{3}{2}}^-$ and ${\frac{5}{2}}^-$ in a comprehensively way and try to assign the $P_{cs}(4338)$ and $P_{cs}(4459)$ in the scenario of pentaquark states consistently.

 The article is arranged as follows:   we obtain the QCD sum rules for the masses and pole residues of  the hidden-charm pentaquark states with the isospin $I=0$ in Sect.2;  in Sect.3, we present the numerical results and discussions; and Sect.4 is reserved for our
conclusion.

\section{QCD sum rules for  the  $udsc\bar{c}$ pentaquark states}
Firstly, let us write down  the two-point correlation functions $\Pi(p)$, $\Pi_{\mu\nu}(p)$ and $\Pi_{\mu\nu\alpha\beta}(p)$,
\begin{eqnarray}\label{CF-Pi-Pi-Pi}
\Pi(p)&=&i\int d^4x e^{ip \cdot x} \langle0|T\left\{J(x)\bar{J}(0)\right\}|0\rangle \, ,\nonumber\\
\Pi_{\mu\nu}(p)&=&i\int d^4x e^{ip \cdot x} \langle0|T\left\{J_{\mu}(x)\bar{J}_{\nu}(0)\right\}|0\rangle \, ,\nonumber\\
\Pi_{\mu\nu\alpha\beta}(p)&=&i\int d^4x e^{ip \cdot x} \langle0|T\left\{J_{\mu\nu}(x)\bar{J}_{\alpha\beta}(0)\right\}|0\rangle \, ,
\end{eqnarray}
where the currents
 \begin{eqnarray}
 J(x)&=&J^1(x)\, , \, J^2(x)\, , \, J^3(x)\, , \, J^4(x)\, , \nonumber\\
 J_\mu(x)&=&J_\mu^1(x)\, , \, J_\mu^2(x)\, , \, J_\mu^3(x)\, , \, J_\mu^4(x)\, , \, J_\mu^5(x)\, , \nonumber\\
 J_{\mu\nu}(x)&=&J_{\mu,\nu}^1(x)\, , \, J_{\mu,\nu}^2(x)\, ,
 \end{eqnarray}
 with
\begin{eqnarray}\label{Current-12}
 J^1(x)&=&\varepsilon^{ila} \varepsilon^{ijk}\varepsilon^{lmn}  u^T_j(x) C\gamma_5 d_k(x)\,s^T_m(x) C\gamma_5 c_n(x)\,  C\bar{c}^{T}_{a}(x) \, , \nonumber\\
J^2(x)&=&\varepsilon^{ila} \varepsilon^{ijk}\varepsilon^{lmn}  u^T_j(x) C\gamma_5 d_k(x)\,s^T_m(x) C\gamma_\mu c_n(x)\,\gamma_5 \gamma^\mu C\bar{c}^{T}_{a}(x) \, ,\nonumber\\
 J^{3}(x)&=&\frac{\varepsilon^{ila} \varepsilon^{ijk}\varepsilon^{lmn}}{\sqrt{2}} \left[u^T_j(x) C\gamma_\mu s_k(x)d^T_m(x) C\gamma^\mu c_n(x) -d^T_j(x) C\gamma_\mu s_k(x)u^T_m(x) C\gamma^\mu c_n(x) \right] C\bar{c}^{T}_{a}(x) \, , \nonumber\\
J^{4}(x)&=&\frac{\varepsilon^{ila} \varepsilon^{ijk}\varepsilon^{lmn}}{\sqrt{2}} \left[u^T_j(x) C\gamma_\mu s_k(x) d^T_m(x) C\gamma_5 c_n(x)-d^T_j(x) C\gamma_\mu s_k(x) u^T_m(x) C\gamma_5 c_n(x)\right] \gamma_5 \gamma^\mu  C\bar{c}^{T}_{a}(x) \, ,\nonumber\\
   \end{eqnarray}
for the isospin-spin $(I,J)=(0,\frac{1}{2})$,
\begin{eqnarray}\label{Current-32}
J^1_\mu(x)&=&\varepsilon^{ila} \varepsilon^{ijk}\varepsilon^{lmn}  u^T_j(x) C\gamma_5 d_k(x)\,s^T_m(x) C\gamma_\mu c_n(x) C\bar{c}^{T}_{a}(x) \, ,\nonumber\\
J^{2}_{\mu}(x)&=&\frac{\varepsilon^{ila} \varepsilon^{ijk}\varepsilon^{lmn}}{\sqrt{2}} \left[ u^T_j(x) C\gamma_5 s_k(x) d^T_m(x) C\gamma_\mu c_n(x)-d^T_j(x) C\gamma_5 s_k(x) u^T_m(x) C\gamma_\mu c_n(x)\right]    C\bar{c}^{T}_{a}(x) \, ,  \nonumber\\
J^{3}_{\mu}(x)&=&\frac{\varepsilon^{ila} \varepsilon^{ijk}\varepsilon^{lmn}}{\sqrt{2}} \left[ u^T_j(x) C\gamma_\mu s_k(x) d^T_m(x) C\gamma_5 c_n(x)-d^T_j(x) C\gamma_\mu s_k(x) u^T_m(x) C\gamma_5 c_n(x)\right]    C\bar{c}^{T}_{a}(x) \, ,  \nonumber\\
 J^{4}_{\mu}(x)&=&\frac{\varepsilon^{ila} \varepsilon^{ijk}\varepsilon^{lmn}}{\sqrt{2}} \left[u^T_j(x) C\gamma_\mu s_k(x)d^T_m(x) C\gamma_\alpha c_n(x) -d^T_j(x) C\gamma_\mu s_k(x)u^T_m(x) C\gamma_\alpha c_n(x) \right] \gamma_5\gamma^\alpha C\bar{c}^{T}_{a}(x) \, ,\nonumber\\
J^{5}_{\mu}(x)&=&\frac{\varepsilon^{ila} \varepsilon^{ijk}\varepsilon^{lmn}}{\sqrt{2}} \left[ u^T_j(x) C\gamma_\alpha s_k(x)d^T_m(x) C\gamma_\mu c_n(x) -d^T_j(x) C\gamma_\alpha s_k(x)u^T_m(x) C\gamma_\mu c_n(x) \right] \gamma_5\gamma^\alpha C\bar{c}^{T}_{a}(x) \, , \nonumber\\
\end{eqnarray}
for the isospin-spin $(I,J)=(0,\frac{3}{2})$,
\begin{eqnarray}\label{Current-52}
J^1_{\mu\nu}(x)&=&\frac{\varepsilon^{ila} \varepsilon^{ijk}\varepsilon^{lmn} }{\sqrt{2}} u^T_j(x) C\gamma_5 d_k(x)\left[s^T_m(x) C\gamma_\mu c_n(x)\, \gamma_5\gamma_{\nu}C\bar{c}^{T}_{a}(x)+s^T_m(x) C\gamma_\nu c_n(x)\,\gamma_5 \gamma_{\mu}C\bar{c}^{T}_{a}(x)\right] \, ,\nonumber\\
J^2_{\mu\nu}(x)&=&\frac{\varepsilon^{ila} \varepsilon^{ijk}\varepsilon^{lmn}}{2} \left[ u^T_j(x) C\gamma_\mu s_k(x)d^T_m(x) C\gamma_\nu c_n(x)-d^T_j(x) C\gamma_\mu s_k(x)u^T_m(x) C\gamma_\nu c_n(x) \right]C\bar{c}^{T}_{a}(x)    \nonumber\\
 &&+\left( \mu\leftrightarrow\nu\right)\, ,  \nonumber\\
\end{eqnarray}
for the isospin-spin $(I,J)=(0,\frac{5}{2})$,
where the $i$, $j$, $k$, $l$, $m$, $n$ and $a$ are color indices, the $C$ is the charge conjugation matrix. We adopt the current $J^1(x)$ and corresponding analysis  in Ref.\cite{WangZG-Pcs4459-333} directly, and construct  other currents according to
the routine shown in Refs.\cite{WangZG-Review,WangZG-EPJC-1509-12,
 WangZG-NPB-1512-32}. We study the mass spectrum of the hidden-charm pentaquark states with the isospin $I=0$ as the $P_{cs}(4459)$ and $P_{cs}(4338)$ were observed in the $J/\psi \Lambda$  invariant mass distribution.

In those currents, there are diquarks $\varepsilon^{ijk}u^T_jC\gamma_5d_k$, $\varepsilon^{ijk}q^T_jC\gamma_{5}s_k$, $\varepsilon^{ijk}q^T_jC\gamma_{\mu}s_k$, $\varepsilon^{ijk}q^T_jC\gamma_5c_k$, $\varepsilon^{ijk}q^T_jC\gamma_{\mu}c_k$, $\varepsilon^{ijk}s^T_jC\gamma_5c_k$, $\varepsilon^{ijk}s^T_jC\gamma_{\mu}c_k$ with $q=u$, $d$. We take  the $S_L$ and $S_H$
to represent  the spins of the light  and heavy diquarks respectively, the  $\varepsilon^{ijk}u^T_jC\gamma_5d_k$, $\varepsilon^{ijk}q^T_jC\gamma_{5}s_k$ and  $\varepsilon^{ijk}q^T_jC\gamma_{\mu}s_k$ have the spins $S_L=0$, $0$ and $1$, respectively, the $\varepsilon^{ijk}q^T_jC\gamma_5c_k$, $\varepsilon^{ijk}s^T_jC\gamma_5c_k$, $\varepsilon^{ijk}q^T_jC\gamma_{\mu}c_k$ and   $\varepsilon^{ijk}s^T_jC\gamma_{\mu}c_k$ have the spins $S_H=0$, $0$, $1$ and $1$, respectively. Then a light diquark and  a heavy diquark form a  tetraquark in the color triplet $\mathbf{3}$ with  angular momentum $\vec{J}_{LH}=\vec{S}_L+\vec{S}_H$, which has the values $J_{LH}=0$, $1$ or $2$.
The operator $C\bar{c}_a^T$ has the spin-parity $J^P={\frac{1}{2}}^-$,
while the  operator $\gamma_5\gamma_{\mu}C\bar{c}_a^T$ has the spin-parity $J^P={\frac{3}{2}}^-$. The total angular momentums   are $\vec{J}=\vec{J}_{LH}+\vec{J}_{\bar{c}}$ with the values $J=\frac{1}{2}$, $\frac{3}{2}$ or $\frac{5}{2}$, which are shown explicitly in Table \ref{current-pentaQ}.

\begin{table}
\begin{center}
\begin{tabular}{|c|c|c|c|c|c|c|c|c|}\hline\hline
$[qq][qc]\bar{c}$ ($S_L$, $S_H$, $J_{LH}$, $J$)                   & $J^{P}$              & Currents              \\ \hline

$[ud][sc]\bar{c}$ ($0$, $0$, $0$, $\frac{1}{2}$)                       &${\frac{1}{2}}^{-}$  &$J^1(x)$              \\

$[ud][sc]\bar{c}$ ($0$, $1$, $1$, $\frac{1}{2}$)                       &${\frac{1}{2}}^{-}$  &$J^2(x)$              \\

$[us][dc]\bar{c}-[ds][uc]\bar{c}$ ($1$, $1$, $0$, $\frac{1}{2}$)       &${\frac{1}{2}}^{-}$  &$J^3(x)$              \\

$[us][dc]\bar{c}-[ds][uc]\bar{c}$ ($1$, $0$, $1$, $\frac{1}{2}$)       &${\frac{1}{2}}^{-}$  &$J^4(x)$             \\

$[ud][sc]\bar{c}$ ($0$, $1$, $1$, $\frac{3}{2}$)                       &${\frac{3}{2}}^{-}$  &$J^1_\mu(x)$              \\

$[us][dc]\bar{c}-[ds][uc]\bar{c}$ ($0$, $1$, $1$, $\frac{3}{2}$)       &${\frac{3}{2}}^{-}$  &$J^2_\mu(x)$          \\

$[us][dc]\bar{c}-[ds][uc]\bar{c}$ ($1$, $0$, $1$, $\frac{3}{2}$)       &${\frac{3}{2}}^{-}$  &$J^3_\mu(x)$          \\

$[us][dc]\bar{c}-[ds][uc]\bar{c}$ ($1$, $1$, $2$, $\frac{3}{2}$)${}_4$ &${\frac{3}{2}}^{-}$  &$J^4_\mu(x)$          \\

$[us][dc]\bar{c}-[ds][uc]\bar{c}$ ($1$, $1$, $2$, $\frac{3}{2}$)${}_5$ &${\frac{3}{2}}^{-}$  &$J^5_\mu(x)$          \\

$[ud][sc]\bar{c}$ ($0$, $1$, $1$, $\frac{5}{2}$)                       &${\frac{5}{2}}^{-}$  &$J^1_{\mu\nu}(x)$     \\

$[us][dc]\bar{c}-[ds][uc]\bar{c}$ ($1$, $1$, $2$, $\frac{5}{2}$)       &${\frac{5}{2}}^{-}$  &$J^2_{\mu\nu}(x)$           \\ \hline\hline
\end{tabular}
\end{center}
\caption{ The quark structures and spin-parity of the  currents.  }\label{current-pentaQ}
\end{table}

The currents $J(x)$, $J_\mu(x)$ and $J_{\mu\nu}(x)$ have the spin-parity
$J^P={\frac{1}{2}}^-$, ${\frac{3}{2}}^-$ and ${\frac{5}{2}}^-$, respectively, and
couple potentially to the hidden-charm pentaquark states (P) with negative  and positive parity,
\begin{eqnarray}\label{Coupling12}
\langle 0| J (0)|P_{\frac{1}{2}}^{-}(p)\rangle &=&\lambda^{-}_{\frac{1}{2}} U^{-}(p,s) \, , \nonumber \\
\langle 0| J (0)|P_{\frac{1}{2}}^{+}(p)\rangle &=&\lambda^{+}_{\frac{1}{2}} i\gamma_5 U^{+}(p,s) \, ,
\end{eqnarray}
\begin{eqnarray}
\langle 0| J_{\mu} (0)|P_{\frac{3}{2}}^{-}(p)\rangle &=&\lambda^{-}_{\frac{3}{2}} U^{-}_\mu(p,s) \, ,  \nonumber \\
\langle 0| J_{\mu} (0)|P_{\frac{3}{2}}^{+}(p)\rangle &=&\lambda^{+}_{\frac{3}{2}}i\gamma_5 U^{+}_\mu(p,s) \, ,  \nonumber \\
\langle 0| J_{\mu} (0)|P_{\frac{1}{2}}^{+}(p)\rangle &=&f^{+}_{\frac{1}{2}}p_\mu U^{+}(p,s) \, , \nonumber \\
\langle 0| J_{\mu} (0)|P_{\frac{1}{2}}^{-}(p)\rangle &=&f^{-}_{\frac{1}{2}}p_\mu i\gamma_5 U^{-}(p,s) \, ,
\end{eqnarray}
\begin{eqnarray}\label{Coupling52}
\langle 0| J_{\mu\nu} (0)|P_{\frac{5}{2}}^{-}(p)\rangle &=&\sqrt{2}\lambda^{-}_{\frac{5}{2}} U^{-}_{\mu\nu}(p,s) \, ,\nonumber\\
\langle 0| J_{\mu\nu} (0)|P_{\frac{5}{2}}^{+}(p)\rangle &=&\sqrt{2}\lambda^{+}_{\frac{5}{2}}i\gamma_5 U^{+}_{\mu\nu}(p,s) \, ,\nonumber\\
\langle 0| J_{\mu\nu} (0)|P_{\frac{3}{2}}^{+}(p)\rangle &=&f^{+}_{\frac{3}{2}} \left[p_\mu U^{+}_{\nu}(p,s)+p_\nu U^{+}_{\mu}(p,s)\right] \, , \nonumber\\
\langle 0| J_{\mu\nu} (0)|P_{\frac{3}{2}}^{-}(p)\rangle &=&f^{-}_{\frac{3}{2}}i\gamma_5 \left[p_\mu U^{-}_{\nu}(p,s)+p_\nu U^{-}_{\mu}(p,s)\right] \, , \nonumber\\
\langle 0| J_{\mu\nu} (0)|P_{\frac{1}{2}}^{-}(p)\rangle &=&g^{-}_{\frac{1}{2}}p_\mu p_\nu U^{-}(p,s) \, , \nonumber\\
\langle 0| J_{\mu\nu} (0)|P_{\frac{1}{2}}^{+}(p)\rangle &=&g^{+}_{\frac{1}{2}}p_\mu p_\nu i\gamma_5 U^{+}(p,s) \, ,
\end{eqnarray}
where the superscripts $\pm$ represent  the  parity, the subscripts $\frac{1}{2}$, $\frac{3}{2}$ and $\frac{5}{2}$ represent  the spins,     the $\lambda$, $f$ and $g$ are the pole residues,  because   multiplying $i \gamma_{5}$ to the currents   $J(x)$, $J_\mu(x)$ and $J_{\mu\nu}(x)$ changes their parity.
The spinors $U^\pm(p,s)$ satisfy the Dirac equations  $(\not\!\!p-M_{\pm})U^{\pm}(p)=0$, while the spinors $U^{\pm}_\mu(p,s)$ and $U^{\pm}_{\mu\nu}(p,s)$ satisfy the Rarita-Schwinger equations $(\not\!\!p-M_{\pm})U^{\pm}_\mu(p)=0$ and $(\not\!\!p-M_{\pm})U^{\pm}_{\mu\nu}(p)=0$,  and the relations $\gamma^\mu U^{\pm}_\mu(p,s)=0$,
$p^\mu U^{\pm}_\mu(p,s)=0$, $\gamma^\mu U^{\pm}_{\mu\nu}(p,s)=0$,
$p^\mu U^{\pm}_{\mu\nu}(p,s)=0$, $ U^{\pm}_{\mu\nu}(p,s)= U^{\pm}_{\nu\mu}(p,s)$, respectively \cite{WangZG-Review,Wang1508-EPJC}.

At the hadron  side, we insert  a complete set  of intermediate hidden-charm pentaquark states with the same quantum numbers as the currents  $J(x)$, $i\gamma_5 J(x)$, $J_{\mu}(x)$, $i\gamma_5 J_{\mu}(x)$, $J_{\mu\nu}(x)$  and $i\gamma_5 J_{\mu\nu}(x)$ into the correlation functions
$\Pi(p)$, $\Pi_{\mu\nu}(p)$ and $\Pi_{\mu\nu\alpha\beta}(p)$ to obtain the hadronic representation
\cite{SVZ79-1,SVZ79-2,PRT85},  isolate the  lowest  states, and obtain the results:
\begin{eqnarray}\label{CF-Hadron-12}
\Pi(p) & = & {\lambda^{-}_{\frac{1}{2}}}^2  {\!\not\!{p}+ M_{-} \over M_{-}^{2}-p^{2}  }+  {\lambda^{+}_{\frac{1}{2}}}^2  {\!\not\!{p}- M_{+} \over M_{+}^{2}-p^{2}  } +\cdots  \, ,\nonumber\\
&=&\Pi_{\frac{1}{2}}^1(p^2)\!\not\!{p}+\Pi_{\frac{1}{2}}^0(p^2)\, ,
 \end{eqnarray}
\begin{eqnarray}\label{CF-Hadron-32}
 \Pi_{\mu\nu}(p) & = & {\lambda^{-}_{\frac{3}{2}}}^2  {\!\not\!{p}+ M_{-} \over M_{-}^{2}-p^{2}  } \left(- g_{\mu\nu}+\frac{\gamma_\mu\gamma_\nu}{3}+\frac{2p_\mu p_\nu}{3p^2}-\frac{p_\mu\gamma_\nu-p_\nu \gamma_\mu}{3\sqrt{p^2}}
\right)\nonumber\\
&&+  {\lambda^{+}_{\frac{3}{2}}}^2  {\!\not\!{p}- M_{+} \over M_{+}^{2}-p^{2}  } \left(- g_{\mu\nu}+\frac{\gamma_\mu\gamma_\nu}{3}+\frac{2p_\mu p_\nu}{3p^2}-\frac{p_\mu\gamma_\nu-p_\nu \gamma_\mu}{3\sqrt{p^2}}
\right)   \nonumber \\
& &+ {f^{+}_{\frac{1}{2}}}^2  {\!\not\!{p}+ M_{+} \over M_{+}^{2}-p^{2}  } p_\mu p_\nu+  {f^{-}_{\frac{1}{2}}}^2  {\!\not\!{p}- M_{-} \over M_{-}^{2}-p^{2}  } p_\mu p_\nu  +\cdots  \, , \nonumber\\
&=&\left[\Pi_{\frac{3}{2}}^1(p^2)\!\not\!{p}+\Pi_{\frac{3}{2}}^0(p^2)\right]\left(- g_{\mu\nu}\right)+\cdots\, ,
\end{eqnarray}
\begin{eqnarray}\label{CF-Hadron-52}
\Pi_{\mu\nu\alpha\beta}(p) & = &2{\lambda^{-}_{\frac{5}{2}}}^2  {\!\not\!{p}+ M_{-} \over M_{-}^{2}-p^{2}  } \left[\frac{ \widetilde{g}_{\mu\alpha}\widetilde{g}_{\nu\beta}+\widetilde{g}_{\mu\beta}\widetilde{g}_{\nu\alpha}}{2}-\frac{\widetilde{g}_{\mu\nu}\widetilde{g}_{\alpha\beta}}{5}-\frac{1}{10}\left( \gamma_{\mu}\gamma_{\alpha}+\frac{\gamma_{\mu}p_{\alpha}-\gamma_{\alpha}p_{\mu}}{\sqrt{p^2}}-\frac{p_{\mu}p_{\alpha}}{p^2}\right)\widetilde{g}_{\nu\beta}\right.\nonumber\\
&&\left.-\frac{1}{10}\left( \gamma_{\nu}\gamma_{\alpha}+\frac{\gamma_{\nu}p_{\alpha}-\gamma_{\alpha}p_{\nu}}{\sqrt{p^2}}-\frac{p_{\nu}p_{\alpha}}{p^2}\right)\widetilde{g}_{\mu\beta}
+\cdots\right]\nonumber\\
&&+  2 {\lambda^{+}_{\frac{5}{2}}}^2  {\!\not\!{p}- M_{+} \over M_{+}^{2}-p^{2}  } \left[\frac{ \widetilde{g}_{\mu\alpha}\widetilde{g}_{\nu\beta}+\widetilde{g}_{\mu\beta}\widetilde{g}_{\nu\alpha}}{2}
-\frac{\widetilde{g}_{\mu\nu}\widetilde{g}_{\alpha\beta}}{5}-\frac{1}{10}\left( \gamma_{\mu}\gamma_{\alpha}+\frac{\gamma_{\mu}p_{\alpha}-\gamma_{\alpha}p_{\mu}}{\sqrt{p^2}}-\frac{p_{\mu}p_{\alpha}}{p^2}\right)\widetilde{g}_{\nu\beta}\right.\nonumber\\
&&\left.
-\frac{1}{10}\left( \gamma_{\nu}\gamma_{\alpha}+\frac{\gamma_{\nu}p_{\alpha}-\gamma_{\alpha}p_{\nu}}{\sqrt{p^2}}-\frac{p_{\nu}p_{\alpha}}{p^2}\right)\widetilde{g}_{\mu\beta}
 +\cdots\right]   \nonumber\\
 && +{f^{+}_{\frac{3}{2}}}^2  {\!\not\!{p}+ M_{+} \over M_{+}^{2}-p^{2}  } \left[ p_\mu p_\alpha \left(- g_{\nu\beta}+\frac{\gamma_\nu\gamma_\beta}{3}+\frac{2p_\nu p_\beta}{3p^2}-\frac{p_\nu\gamma_\beta-p_\beta \gamma_\nu}{3\sqrt{p^2}}
\right)+\cdots \right]\nonumber\\
&&+  {f^{-}_{\frac{3}{2}}}^2  {\!\not\!{p}- M_{-} \over M_{-}^{2}-p^{2}  } \left[ p_\mu p_\alpha \left(- g_{\nu\beta}+\frac{\gamma_\nu\gamma_\beta}{3}+\frac{2p_\nu p_\beta}{3p^2}-\frac{p_\nu\gamma_\beta-p_\beta \gamma_\nu}{3\sqrt{p^2}}
\right)+\cdots \right]   \nonumber \\
& &+ {g^{-}_{\frac{1}{2}}}^2  {\!\not\!{p}+ M_{-} \over M_{-}^{2}-p^{2}  } p_\mu p_\nu p_\alpha p_\beta+  {g^{+}_{\frac{1}{2}}}^2  {\!\not\!{p}- M_{+} \over M_{+}^{2}-p^{2}  } p_\mu p_\nu p_\alpha p_\beta  +\cdots \, , \nonumber\\
& = & \left[\Pi_{\frac{5}{2}}^1(p^2)\!\not\!{p}+\Pi_{\frac{5}{2}}^0(p^2)\right]\left( g_{\mu\alpha}g_{\nu\beta}+g_{\mu\beta}g_{\nu\alpha}\right)  +\cdots \, ,
 \end{eqnarray}
where $\widetilde{g}_{\mu\nu}=g_{\mu\nu}-\frac{p_{\mu}p_{\nu}}{p^2}$. We prefer to the components $\Pi_{\frac{1}{2}}^1(p^2)$, $\Pi_{\frac{1}{2}}^0(p^2)$, $\Pi_{\frac{3}{2}}^1(p^2)$, $\Pi_{\frac{3}{2}}^0(p^2)$, $\Pi_{\frac{5}{2}}^1(p^2)$ and  $\Pi_{\frac{5}{2}}^0(p^2)$ to avoid possible contaminations from other pentaquark states with different spins.

Then we obtain the spectral densities  through  dispersion relation,
\begin{eqnarray}
\frac{{\rm Im}\Pi^1_j(s)}{\pi}&=& \lambda_{-}^2 \delta\left(s-M_{-}^2\right)+\lambda_{+}^2 \delta\left(s-M_{+}^2\right) =\, \rho^1_{H}(s) \, , \\
\frac{{\rm Im}\Pi^0_j(s)}{\pi}&=&M_{-}\lambda_{-}^2 \delta\left(s-M_{-}^2\right)-M_{+}\lambda_{+}^2 \delta\left(s-M_{+}^2\right)
=\rho^0_{H}(s) \, ,
\end{eqnarray}
where $j=\frac{1}{2}$, $\frac{3}{2}$, $\frac{5}{2}$, we introduce the subscript $H$ to represent  the hadron side,
then we introduce the  weight functions $\sqrt{s}\exp\left(-\frac{s}{T^2}\right)$ and $\exp\left(-\frac{s}{T^2}\right)$ to obtain the QCD sum rules
at the hadron side,
\begin{eqnarray}
\int_{4m_c^2}^{s_0}ds \left[\sqrt{s}\,\rho^1_{H}(s)+\rho^0_{H}(s)\right]\exp\left( -\frac{s}{T^2}\right)
&=&2M_{-}\lambda_{-}^2\exp\left( -\frac{M_{-}^2}{T^2}\right) \, ,
\end{eqnarray}
\begin{eqnarray}
\int_{4m_c^2}^{s^\prime_0}ds \left[\sqrt{s}\,\rho^1_{H}(s)-\rho^0_{H}(s)\right]\exp\left( -\frac{s}{T^2}\right)
&=&2M_{+}\lambda_{+}^2\exp\left( -\frac{M_{+}^2}{T^2}\right) \, ,
\end{eqnarray}
where the $s_0$ and $s_0^\prime$ are the continuum threshold parameters, and the $T^2$ is the Borel parameter. Thus we  distinguish  the  contributions  of the hidden-charm pentaquark states with negative and positive parity unambiguously.

In the QCD sum rules, we choose the local four-quark or five-quark currents, while the traditional mesons and baryons are spatial extended objects and have mean spatial sizes $\sqrt{\langle r^2\rangle} \neq 0$, for example, $\sqrt{\langle r^2\rangle_{\pi^+}}=0.659 \pm 0.004\,\rm{fm}$, $\sqrt{\langle r^2\rangle_{K^+}}=0.560\pm 0.031\,\rm{fm}$, $\sqrt{\langle r^2\rangle_{E,p}}=0.8409 \pm0.0004\,\rm{fm}$, $\sqrt{\langle r^2\rangle_{M,p}}=0.851\pm 0.026\,\rm{fm}$
 from the Particle Data Group \cite{PDG},
where the subscripts $E$ and $M$ stand for the electric  and magnetic radii, respectively, and
$\sqrt{\langle r^2\rangle_{J/\psi}}=0.41\,\rm{fm}$ from the screened potential  model \cite{X3872-charmonium-KTChao-2009-PRD}.
We obtain excellent QCD sum rules for the traditional  mesons and baryons \cite{SVZ79-1,SVZ79-2,PRT85,Ioffe-baryon-mass,Ioffe-baryon-magnetic}, and the QCD sum rules work well for the spatial sizes $\sqrt{\langle r^2\rangle} < 1 \,\rm{fm}$ at least.
In the dynamical diquark model of the multiquarks with the Born-Oppenheimer potentials calculated numerically on the lattice, the average sizes $\langle r\rangle < 1 \,\rm{fm}$ for all the hidden-charm pentaquark states except for only one case for the excited 2D states \cite{Penta-radii-Lebed-JHEP}. If the exotic states have the average spatial sizes as that of the typical mesons and baryons, such as the $\pi$, $K$, $J/\psi$, $p$, we expect that the QCD sum rules also work well for the exotic states, in fact, the QCD sum rules have given many successful descriptions \cite{WangZG-Review,Nielsen-review}.

If we perform Fierz transformations for the interpolating currents shown in Eqs.\eqref{Current-12}-\eqref{Current-52} both in the Dirac spinor and color space, just like what we have done for the four-quark currents \cite{WangZG-two-part-IJPMA-2020}, we obtain a superposition of a series of  color singlet-singlet type currents, for example,
\begin{eqnarray}\label{Fierz-1}
\tilde{J}^1&=& 2iS_{ud} c\, \bar{c} i\gamma_5 s
-2S_{ud} \gamma_5 c\, \bar{c} s +2S_{ud} \gamma_{\alpha} c \,\bar{c} \gamma^{\alpha}\gamma_5 s
+2S_{ud} \gamma_{\alpha}\gamma_5 c\, \bar{c} \gamma^{\alpha} s\nonumber\\
&&+S_{ud} \sigma_{\alpha\beta}\gamma_5 c \,\bar{c} \sigma^{\alpha\beta} s  -2iS_{ud} s \,\bar{c} i\gamma_5 c+2S_{ud} \gamma_5 s \,\bar{c} c
-2S_{ud} \gamma_{\alpha} s \,\bar{c} \gamma^{\alpha}\gamma_5 c\nonumber\\
&&-2S_{ud} \gamma_{\alpha}\gamma_5 s\, \bar{c} \gamma^{\alpha} c
-S_{ud} \sigma_{\alpha\beta}\gamma_5 s \,\bar{c} \sigma^{\alpha\beta} c\, ,
\end{eqnarray}

\begin{eqnarray}\label{Fierz-2}
\tilde{J}^2&=& -4S_{ud} \gamma_5 c \, \bar{c} s
+S_{ud} \gamma_5\sigma_{\alpha\mu} c \,\bar{c} \sigma^{\mu\alpha} s -4iS_{ud} c\, \bar{c} i\gamma_5 s
-S_{ud} \sigma_{\alpha\mu} c \,\bar{c} \gamma_5\sigma^{\mu\alpha} s\nonumber\\
&&-2S_{ud} \gamma_{\alpha} c \,\bar{c} \gamma^{\alpha}\gamma_5 s
-2S_{ud} \gamma_5\gamma_{\alpha} c \,\bar{c} \gamma^{\alpha} s-4S_{ud} \gamma_5 s \,\bar{c} c
+S_{ud} \gamma_5\sigma_{\alpha\mu} s\, \bar{c} \sigma^{\mu\alpha} c\nonumber\\
&&-4iS_{ud} s\, \bar{c} i\gamma_5 c
-S_{ud} \sigma_{\alpha\mu} s\, \bar{c} \gamma_5\sigma^{\mu\alpha} c-2S_{ud} \gamma_{\alpha} s \,\bar{c} \gamma^{\alpha}\gamma_5 c
-2S_{ud} \gamma_5\gamma_{\alpha} s \,\bar{c} \gamma^{\alpha} c\, ,
\end{eqnarray}

\begin{eqnarray}\label{Fierz-3}
\tilde{J}^3&=&iA_{us,\mu} \gamma_5\gamma^{\mu} c \,\bar{c} i\gamma_5 d -iA_{ds,\mu} \gamma_5\gamma^{\mu} c\, \bar{c} i\gamma_5 u-A_{us,\mu} \gamma^{\mu} c\, \bar{c} d
+A_{ds,\mu} \gamma^{\mu} c\, \bar{c} u\nonumber\\
&&-A_{us,\mu} \gamma_5 c\, \bar{c} \gamma^{\mu}\gamma_5 d
+A_{ds,\mu} \gamma_5 c\, \bar{c} \gamma^{\mu}\gamma_5 u+iA_{us,\mu} \gamma_5 \sigma^{\alpha\mu}c\, \bar{c} \gamma_{\alpha}\gamma_5 d -iA_{ds,\mu} \gamma_5 \sigma^{\alpha\mu}c\, \bar{c} \gamma_{\alpha}\gamma_5 u\nonumber\\
&&+A_{us,\mu} c\, \bar{c} \gamma^{\mu} d
-A_{ds,\mu} c\, \bar{c} \gamma^{\mu} u-iA_{us,\mu} \sigma^{\alpha\mu} c\, \bar{c} \gamma_{\alpha} d
+iA_{ds,\mu} \sigma^{\alpha\mu} c\, \bar{c} \gamma_{\alpha} u\nonumber\\
&&+iA_{us,\mu} \gamma_{\alpha} c\, \bar{c} \sigma^{\alpha\mu} d
-iA_{ds,\mu} \gamma_{\alpha} c\, \bar{c} \sigma^{\alpha\mu} u-iA_{us,\mu} \gamma_5\gamma_{\alpha} c\, \bar{c} \gamma_5\sigma^{\mu\alpha} d +iA_{ds,\mu} \gamma_5\gamma_{\alpha} c\, \bar{c} \gamma_5\sigma^{\mu\alpha} u\nonumber\\
&&+iA_{us,\mu} \gamma_5\gamma^{\mu} d\, \bar{c} i\gamma_5 c
-iA_{ds,\mu} \gamma_5\gamma^{\mu} u\, \bar{c} i\gamma_5 c-A_{us,\mu} \gamma^{\mu} d\, \bar{c} c
+A_{ds,\mu} \gamma^{\mu} u\, \bar{c} c\nonumber\\
&&-A_{us,\mu} \gamma_5 d\, \bar{c} \gamma^{\mu}\gamma_5 c
+A_{ds,\mu} \gamma_5 u\, \bar{c} \gamma^{\mu}\gamma_5 c+iA_{us,\mu} \gamma_5\sigma^{\alpha\mu} d\, \bar{c} \gamma_{\alpha}\gamma_5 c -iA_{ds,\mu} \gamma_5\sigma^{\alpha\mu} u\, \bar{c} \gamma_{\alpha}\gamma_5 c\nonumber\\
&&+A_{us,\mu} d\, \bar{c} \gamma^{\mu} c
-A_{ds,\mu} u\, \bar{c} \gamma^{\mu} c-iA_{us,\mu} \sigma^{\alpha\mu} d\, \bar{c} \gamma_{\alpha} c
+iA_{ds,\mu} \sigma^{\alpha\mu} u\, \bar{c} \gamma_{\alpha} c+iA_{us,\mu} \gamma_{\alpha} d\, \bar{c} \sigma^{\alpha\mu} c\nonumber\\
&&
-iA_{ds,\mu} \gamma_{\alpha} u\, \bar{c} \sigma^{\alpha\mu} c-iA_{us,\mu} \gamma_5\gamma_{\alpha} d\, \bar{c} \gamma_5\sigma^{\mu\alpha} c +iA_{ds,\mu} \gamma_5\gamma_{\alpha} u\, \bar{c} \gamma_5\sigma^{\mu\alpha} c\, ,
\end{eqnarray}

\begin{eqnarray}\label{Fierz-4}
\tilde{J}^4&=&A_{us,\mu} c \,\bar{c} \gamma^{\mu} d
-A_{ds,\mu} c \,\bar{c} \gamma^{\mu} u+A_{us,\mu} \gamma_5 c \,\bar{c} \gamma^{\mu}\gamma_5 d
-A_{ds,\mu} \gamma_5 c\, \bar{c} \gamma^{\mu}\gamma_5 u\nonumber\\
&&+A_{us,\mu} \gamma^{\mu} c \,\bar{c} d
-A_{ds,\mu} \gamma^{\mu} c\, \bar{c} u-iA_{us,\mu} \gamma_{\alpha} c \,\bar{c} \sigma^{\mu\alpha} d
+iA_{ds,\mu} \gamma_{\alpha} c\, \bar{c} \sigma^{\mu\alpha} u\nonumber\\
&&-iA_{us,\mu} \gamma^{\mu}\gamma_5 c\, \bar{c} i\gamma_5 d
+iA_{ds,\mu} \gamma^{\mu}\gamma_5 c\, \bar{c} i\gamma_5 u-iA_{us,\mu} \gamma_{\alpha}\gamma_5 c\, \bar{c} \gamma_5\sigma^{\mu\alpha} d
+iA_{ds,\mu} \gamma_{\alpha}\gamma_5 c\bar{c} \gamma_5\sigma^{\mu\alpha} u_m\nonumber\\
&&+iA_{us,\mu} \sigma^{\mu\alpha}\gamma_5 c \bar{c}_m \gamma_5\gamma_{\alpha} d_m
-iA_{ds,\mu} \sigma^{\mu\alpha}\gamma_5 c_k \bar{c}_m \gamma_5\gamma_{\alpha} u-iA_{us,\mu} \sigma^{\mu\alpha} c \,\bar{c} \gamma_{\alpha} d
+iA_{ds,\mu} \sigma^{\mu\alpha} c\, \bar{c} \gamma_{\alpha} u\nonumber\\
&&-A_{us,\mu} d\, \bar{c} \gamma^{\mu} c
+A_{ds,\mu} u \,\bar{c} \gamma^{\mu} c-A_{us,\mu} \gamma_5 d\, \bar{c} \gamma^{\mu}\gamma_5 c
+A_{ds,\mu} \gamma_5 u\, \bar{c} \gamma^{\mu}\gamma_5 c\nonumber\\
&&-A_{us,\mu} \gamma^{\mu} d \,\bar{c} c
+A_{ds,\mu} \gamma^{\mu} u\, \bar{c} c+iA_{us,\mu} \gamma_{\alpha} d \,\bar{c} \sigma^{\mu\alpha} c
-iA_{ds,\mu} \gamma_{\alpha} u\, \bar{c} \sigma^{\mu\alpha} c\nonumber\\
&&+iA_{us,\mu} \gamma^{\mu}\gamma_5 d\, \bar{c} i\gamma_5 c
-iA_{ds,\mu} \gamma^{\mu}\gamma_5 u\, \bar{c}i\gamma_5 c+iA_{us,\mu} \gamma_{\alpha}\gamma_5 d \,\bar{c}\gamma_5\sigma^{\mu\alpha} c
-iA_{ds,\mu} \gamma_{\alpha}\gamma_5 u\, \bar{c} \gamma_5\sigma^{\mu\alpha} c\nonumber\\
&&-iA_{us,\mu} \sigma^{\mu\alpha}\gamma_5 d \,\bar{c}\gamma_5\gamma_{\alpha} c
+iA_{ds,\mu} \sigma^{\mu\alpha}\gamma_5 u\, \bar{c} \gamma_5\gamma_{\alpha} c+iA_{us,\mu} \sigma^{\mu\alpha} d \,\bar{c} \gamma_{\alpha} c
-iA_{ds,\mu} \sigma^{\mu\alpha} u \bar{c} \gamma_{\alpha} c\, ,
\end{eqnarray}
where $\tilde{J}^1=8J^1(x)$, $\tilde{J}^2=4J^2(x)$,
$\tilde{J}^3=4\sqrt{2}J^3(x)$, $\tilde{J}^4=4\sqrt{2}J^4(x)$,
$S_{ud}\Gamma c=\varepsilon^{ijk}u^T_i C\gamma_5 d_j \Gamma c_k$,
$A_{us,\mu}\Gamma c=\varepsilon^{ijk}u^T_i C\gamma_\mu s_j \Gamma c_k$,
$A_{ds,\mu}\Gamma c=\varepsilon^{ijk}d^T_i C\gamma_\mu s_j \Gamma c_k$, and the $\Gamma$ denotes some Dirac $\gamma$-matrixes.

On the other hand, we can interpolate the ground state $\Lambda$ baryon by the following three currents,
\begin{eqnarray}
\eta_1(x)&=&\varepsilon^{ijk}u^T_i(x)C\gamma_5 d_j(x) s_c(x)\, , \nonumber\\
\eta_2(x)&=&\frac{\varepsilon^{ijk}}{\sqrt{2}} \left[s^T_i(x)C\gamma_5 u_j(x) d_c(x)-s^T_i(x)C\gamma_5 d_j(x) u_c(x)\right]\, , \nonumber\\
\eta_3(x)&=&\frac{\varepsilon^{ijk}}{\sqrt{2}} \left[s^T_i(x)C\gamma_\mu u_j(x) \gamma_5\gamma^{\mu}d_c(x)-s^T_i(x)C\gamma_\mu d_j(x) \gamma_5\gamma^{\mu}u_c(x)\right]\, ,
\end{eqnarray}
or their superpositions.
Again, for example, the components $S_{ud} s\, \bar{c} i\gamma_5 c$ and
$A_{us,\mu} \gamma^{\mu}\gamma_5 d\, \bar{c} i\gamma_5 c
-A_{ds,\mu} \gamma^{\mu}\gamma_5 u\, \bar{c}i\gamma_5 c$
couple potentially to the meson-baryon pair $\Lambda \eta_c$, the components
 $S_{ud} s\, \bar{c} \gamma^{\alpha} c$, $A_{us,\mu} d\, \bar{c} \gamma^{\mu} c
-A_{ds,\mu} u\, \bar{c} \gamma^{\mu} c$ and $A_{us,\mu} \gamma_{\alpha}\gamma_5 d \,\bar{c}\gamma_5\sigma^{\mu\alpha} c
-A_{ds,\mu} \gamma_{\alpha}\gamma_5 u\, \bar{c} \gamma_5\sigma^{\mu\alpha} c$ couple potentially to the meson-baryon pair $\Lambda J/\psi$.
 The quantum field theory  does not forbid the couplings between the five-quark currents and baryon-meson scattering states with the average spatial sizes $\sqrt{\langle r^2\rangle} \geq 1 \,\rm{fm}$ if they have the same quantum numbers, also see other components in Eqs.\eqref{Fierz-1}-\eqref{Fierz-4},  however, such couplings are suppressed   as the overlaps of the wave-functions are very small \cite{WangZG-Review,WZG-local-current}.
In other words, local currents couple potentially to the compact exotic states having the average spatial sizes as that of the typical mesons and baryons, not to the two-particle scattering states with average  spatial size $\sqrt{\langle r^2\rangle}\geq 1.0\,\rm{fm}$, which are too large to be interpolated  by the local currents \cite{WangZG-Review,WZG-local-current}.

We  study the contributions of the intermediate meson-baryon scattering states  $\Lambda J/\psi$, $\Lambda \eta_c$, $\Lambda_c \bar{D}_s$, $\Lambda_c \bar{D}^*_s$,  $\cdots$ etc besides the hidden-charm pentaquark states $P_{cs}$ to the  components $\Pi_{j}^{1}(p^2)$ (which corresponds to the traditional QCD sum rules in Eq.\eqref{Traditional-QCDSR-1}) as an example for simplicity,
\begin{eqnarray}\label{Self-Energy}
\Pi_{j}(p^2) &=&-\frac{\widehat{\lambda}_{P}^{2}}{ p^2-\widehat{M}_{P}^2-\Sigma_{\Lambda J/\psi}(p^2)-\Sigma_{\Lambda \eta_c}(p^2)-\Sigma_{\Lambda_c\bar{D}_s}(p^2)+\cdots}+\cdots \, ,
\end{eqnarray}
where
$j=\frac{1}{2}$, $\frac{3}{2}$, $\frac{5}{2}$. We choose the bare quantities $\widehat{\lambda}_{P}$ and $\widehat{M}_{P}$  to absorb the divergences in the self-energies $\Sigma_{\Lambda J/\psi}(p^2)$, $\Sigma_{\Lambda \eta_c}(p^2)$, $\Sigma_{\Lambda_c\bar{D}_s}(p^2)$, etc. The renormalized energies  satisfy  the relation $p^2-M_{P}^2-\overline{\Sigma}_{\Lambda J/\psi}(p^2)-\overline{\Sigma}_{\Lambda \eta_c}(p^2)-\overline{\Sigma}_{\Lambda_c\bar{D}_s}(p^2)+\cdots=0$, where the overlines above the
self-energies denote that the divergent terms have been subtracted. As the pentaquark  states $P_{cs}$ are unstable, the relation should be modified,
$p^2-M_{P}^2-{\rm Re}\overline{\Sigma}_{\Lambda J/\psi}(p^2)-{\rm Re}\overline{\Sigma}_{\Lambda \eta_c}(p^2)-{\rm Re}\overline{\Sigma}_{\Lambda_c\bar{D}_s}(p^2)+\cdots=0$, and
$-{\rm Im}\overline{\Sigma}_{\Lambda J/\psi}(p^2)-{\rm Im}\overline{\Sigma}_{\Lambda \eta_c}(p^2)-{\rm Im}\overline{\Sigma}_{\Lambda_c\bar{D}_s}(p^2)+\cdots=\sqrt{p^2}\Gamma(p^2)$.
The renormalized self-energies  contribute  a finite imaginary part to modify the dispersion relation \cite{WangZG-Landau-PRD-2020},
\begin{eqnarray}\label{Modify-width}
\Pi_{j}^1(p) &=&-\frac{\lambda_{P}^{2}}{ p^2-M_{P}^2+i\sqrt{p^2}\Gamma(p^2)}+\cdots \, .
 \end{eqnarray}

We  take  account of the finite width effect by the  simple replacement of the hadronic spectral density,
\begin{eqnarray}
\lambda^2_{P}\delta \left(s-M^2_{P} \right) &\to& \lambda^2_{P}\frac{1}{\pi}\frac{M_{P}\Gamma_{P}}{(s-M_{P}^2)^2+M_{P}^2\Gamma_{P}^2}\, .
\end{eqnarray}
Then the hadron  sides of  the QCD sum rules undergo the change,
\begin{eqnarray}
\lambda^2_{P}\exp \left(-\frac{M^2_{P}}{T^2} \right) &\to& \lambda^2_{P}\int_{(m_{\Lambda}+m_{\eta_c})^2}^{s_0}
ds\frac{1}{\pi}\frac{M_{P}\Gamma_{P}}{(s-M_{P}^2)^2+M_{P}^2\Gamma_{P}^2}\exp \left(-\frac{s}{T^2} \right)\, , \nonumber\\
&=&C_P^2\,\lambda^2_{P}\exp \left(-\frac{M^2_{P}}{T^2} \right)\, .
\end{eqnarray}
In the case of the current $J^1(x)$, $C_P=0.99$, $0.97$, $0.94$ and  $0.90$ for the widths $\Gamma_P=50\,\rm{MeV}$, $100\,\rm{MeV}$, $200\,\rm{MeV}$ and $300\,\rm{MeV}$, respectively, for  the central values shown in Tables \ref{Borel}-\ref{mass-Pcs}. In fact, the $P_{cs}$ states have the widths about $20\,\rm{MeV}$ \cite{LHCb-Pcs4459-2012,LHCb-Pcs4338}, we  can absorb the numerical factors  $C_P$ into the pole residues safely, the intermediate   meson-baryon loops cannot  affect  the mass $M_{P}$ significantly.

In the QCD sum rules,  we choose the local currents which couple potentially to compact objects, and obtain the color $\bar{\mathbf{3}}\mathbf{3}$-type, $\mathbf{6}\bar{\mathbf{6}}$-type, $\mathbf{1}\mathbf{1}$-type or $\mathbf{8}\mathbf{8}$-type tetraquark states, and  $\bar{\mathbf{3}}\bar{\mathbf{3}}\bar{\mathbf{3}} $-type or $\mathbf{1}\mathbf{1}$-type pentaquark states, although we usually call the $\mathbf{1}\mathbf{1}$-type states as the molecular states.

\begin{figure}
 \centering
 \includegraphics[totalheight=3.0cm,width=15cm]{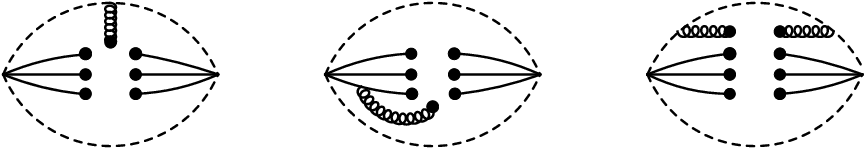}
  \vglue+3mm
 \includegraphics[totalheight=3.0cm,width=15cm]{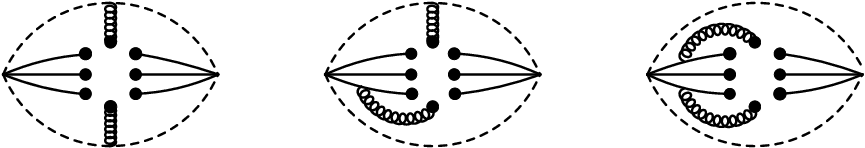}
    \caption{The diagrams contribute  to the  condensates $\langle\bar{q} q\rangle^2\langle\bar{q}g_s\sigma Gq\rangle $, $\langle\bar{q} q\rangle \langle\bar{q}g_s\sigma Gq\rangle^2 $, $\langle \bar{q}q\rangle^3\langle \frac{\alpha_s}{\pi}GG\rangle$, where $q=u$, $d$ or $s$. Other
diagrams obtained by interchanging of the $c$ quark lines (dashed lines) or light quark lines (solid lines) are implied. }\label{Feynman}
\end{figure}

At the QCD side,  we carry out the operator product expansion with the help of the full $u$, $d$, $s$ and $c$ quark propagators,
 \begin{eqnarray}
U/D_{ij}(x)&=& \frac{i\delta_{ij}\!\not\!{x}}{ 2\pi^2x^4}-\frac{\delta_{ij}\langle
\bar{q}q\rangle}{12} -\frac{\delta_{ij}x^2\langle \bar{q}g_s\sigma Gq\rangle}{192} -\frac{ig_sG^{a}_{\alpha\beta}t^a_{ij}(\!\not\!{x}
\sigma^{\alpha\beta}+\sigma^{\alpha\beta} \!\not\!{x})}{32\pi^2x^2} -\frac{\delta_{ij}x^4\langle \bar{q}q \rangle\langle g_s^2 GG\rangle}{27648} \nonumber\\
&&  -\frac{1}{8}\langle\bar{q}_j\sigma^{\mu\nu}q_i \rangle \sigma_{\mu\nu}+\cdots \, ,
\end{eqnarray}
\begin{eqnarray}
S_{ij}(x)&=& \frac{i\delta_{ij}\!\not\!{x}}{ 2\pi^2x^4}
-\frac{\delta_{ij}m_s}{4\pi^2x^2}-\frac{\delta_{ij}\langle
\bar{s}s\rangle}{12} +\frac{i\delta_{ij}\!\not\!{x}m_s
\langle\bar{s}s\rangle}{48}-\frac{\delta_{ij}x^2\langle \bar{s}g_s\sigma Gs\rangle}{192}+\frac{i\delta_{ij}x^2\!\not\!{x} m_s\langle \bar{s}g_s\sigma
 Gs\rangle }{1152}\nonumber\\
&& -\frac{ig_s G^{a}_{\alpha\beta}t^a_{ij}(\!\not\!{x}
\sigma^{\alpha\beta}+\sigma^{\alpha\beta} \!\not\!{x})}{32\pi^2x^2} -\frac{\delta_{ij}x^4\langle \bar{s}s \rangle\langle g_s^2 GG\rangle}{27648}-\frac{1}{8}\langle\bar{s}_j\sigma^{\mu\nu}s_i \rangle \sigma_{\mu\nu}  +\cdots \, ,
\end{eqnarray}
\begin{eqnarray}
C_{ij}(x)&=&\frac{i}{(2\pi)^4}\int d^4k e^{-ik \cdot x} \left\{
\frac{\delta_{ij}}{\!\not\!{k}-m_c}
-\frac{g_sG^n_{\alpha\beta}t^n_{ij}}{4}\frac{\sigma^{\alpha\beta}(\!\not\!{k}+m_c)+(\!\not\!{k}+m_c)
\sigma^{\alpha\beta}}{(k^2-m_c^2)^2}\right.\nonumber\\
&&\left. -\frac{g_s^2 (t^at^b)_{ij} G^a_{\alpha\beta}G^b_{\mu\nu}(f^{\alpha\beta\mu\nu}+f^{\alpha\mu\beta\nu}+f^{\alpha\mu\nu\beta}) }{4(k^2-m_c^2)^5}+\cdots\right\} \, ,\nonumber\\
f^{\alpha\beta\mu\nu}&=&(\!\not\!{k}+m_c)\gamma^\alpha(\!\not\!{k}+m_c)\gamma^\beta(\!\not\!{k}+m_c)\gamma^\mu(\!\not\!{k}+m_c)\gamma^\nu(\!\not\!{k}+m_c)\, ,
\end{eqnarray}
and  $t^n=\frac{\lambda^n}{2}$, the $\lambda^n$ is the Gell-Mann matrix
\cite{PRT85,Pascual-1984,WangHuang3900}.
We introduce  the $\langle\bar{q}_j\sigma_{\mu\nu}q_i \rangle$ and $\langle\bar{s}_j\sigma_{\mu\nu}s_i \rangle$ come from Fierz re-ordering  of the
$\langle q_i \bar{q}_j\rangle$ and $\langle s_i \bar{s}_j\rangle$  to  absorb the gluons  emitted from other quark lines to  extract the mixed condensates   $\langle\bar{q}g_s\sigma G q\rangle$ and $\langle\bar{s}g_s\sigma G s\rangle$, respectively \cite{WangHuang3900}.  Then we compute  all the Feynman diagrams to obtain analytical expressions, and finally obtain the QCD spectral densities through   dispersion relation,
\begin{eqnarray}\label{QCD-rho}
 \rho^1_{QCD}(s) &=&\frac{{\rm Im}\Pi^1_j(s)}{\pi}\, , \nonumber\\
\rho^0_{QCD}(s) &=&\frac{{\rm Im}\Pi^0_j(s)}{\pi}\, ,
\end{eqnarray}
where $j=\frac{1}{2}$, $\frac{3}{2}$, $\frac{5}{2}$.
According to analysis in previous works \cite{WangZG-Pcs4459-333,WangZG-Review,WZG-penta-IJMPA}, we take account of the quark-gluon operators up to dimension $13$ and order $\mathcal{O}( \alpha_s^{k})$ with $k\leq 1$ consistently, then take their vacuum expectations, and   take account of the terms  $\propto m_s$ to account for the light-flavor   $SU(3)$ mass-breaking effects.    The higher dimensional  vacuum condensates, especially the vacuum condensates of dimension 11 and 13, which come from the Feynman  diagrams shown in Fig.\ref{Feynman},  are  associated with the  $\frac{1}{T^2}$, $\frac{1}{T^4}$, $\frac{1}{T^6}$ or $\frac{1}{T^8}$, and manifest themselves at  the small values of the Borel parameter $T^2$ and play an important role in determining the Borel windows \cite{WangZG-Pcs4459-333,WangZG-Review,WZG-penta-IJMPA}.

Now we  match the hadron side with the QCD side of the correlation functions, take the quark-hadron duality below the continuum thresholds, and  obtain  two  QCD sum rules:
\begin{eqnarray}\label{QCDSR}
2M_{-}\lambda_{-}^2\exp\left( -\frac{M_{-}^2}{T^2}\right)&=& \int_{4m_c^2}^{s_0}ds \,\left[\sqrt{s}\rho_{QCD}^1(s)+\rho_{QCD}^{0}(s)\right]\,\exp\left( -\frac{s}{T^2}\right)\,  ,
\end{eqnarray}
\begin{eqnarray}\label{QCDSR-Positive}
2M_{+}\lambda_{+}^2\exp\left( -\frac{M_{+}^2}{T^2}\right)&=& \int_{4m_c^2}^{s^\prime_0}ds \,\left[\sqrt{s}\rho_{QCD}^1(s)-\rho_{QCD}^{0}(s)\right]\,\exp\left( -\frac{s}{T^2}\right)\,  .
\end{eqnarray}

If we set the couplings to the hidden-charm pentaquark states with positive parity to be zero, i.e. $\lambda_{+}=0$, we obtain two traditional QCD sum rules,
\begin{eqnarray}\label{Traditional-QCDSR-1}
\lambda_{-}^2\exp\left( -\frac{M_{-}^2}{T^2}\right)&=&\int_{4m_c^2}^{s_0}ds \,\rho^1_{QCD}(s)\exp\left( -\frac{s}{T^2}\right) \, ,
\end{eqnarray}
\begin{eqnarray}\label{Traditional-QCDSR-0}
M_{-}\lambda_{-}^2\exp\left( -\frac{M_{-}^2}{T^2}\right)&=&\int_{4m_c^2}^{s_0}ds \,\rho^0_{QCD}(s)\exp\left( -\frac{s}{T^2}\right) \, ,
\end{eqnarray}
with respect to the components $\Pi_j^1(p^2)$ and $\Pi^0_j(p^2)$, respectively.
However, such an approximation leads  to contaminations because $\lambda_{+}\neq 0$.

In this work, we adopt the QCD sum rules for the hidden-charm pentaquark states with negative parity, see Eq.\eqref{QCDSR}, and resort to the QCD sum rules for the hidden-charm pentaquark states with positive parity, see Eq.\eqref{QCDSR-Positive}, to estimate the possible contaminations from the hidden-charm pentaquark states with positive parity, if  the two QCD sum rules in Eqs.\eqref{Traditional-QCDSR-1}-\eqref{Traditional-QCDSR-0} are adopted. Now we define a parameter CTM to measure contaminations from the hidden-charm pentaquark states with positive parity,
\begin{eqnarray}
{\rm CTM}&=&\frac{\int_{4m_c^2}^{s_0}ds \,\left[\sqrt{s}\rho_{QCD}^1(s)-\rho_{QCD}^{0}(s)\right]\,\exp\left( -\frac{s}{T^2}\right)}{\int_{4m_c^2}^{s_0}ds \,\left[\sqrt{s}\rho_{QCD}^1(s)+\rho_{QCD}^{0}(s)\right]\,\exp\left( -\frac{s}{T^2}\right)}\, ,
\end{eqnarray}
by setting $s^\prime_0=s_0$.

We differentiate Eq.\eqref{QCDSR} in regard  to  $\frac{1}{T^2}$, then eliminate the pole residues $\lambda_{-}$ and obtain the QCD sum rules for
 the hidden-charm pentaquark masses,
 \begin{eqnarray}
 M^2_{-} &=& \frac{-\int_{4m_c^2}^{s_0}ds \frac{d}{d(1/T^2)}\, \left[\sqrt{s}\rho_{QCD}^1(s)+\rho_{QCD}^{0}(s)\right]\,\exp\left( -\frac{s}{T^2}\right)}{\int_{4m_c^2}^{s_0}ds \, \left[\sqrt{s}\rho_{QCD}^1(s)+\rho_{QCD}^{0}(s)\right]\,\exp\left( -\frac{s}{T^2}\right)}\,  .
\end{eqnarray}

\section{Numerical results and discussions}
We take  the standard values of the  vacuum condensates
$\langle\bar{q}q \rangle=-(0.24\pm 0.01\, \rm{GeV})^3$,  $\langle\bar{s}s \rangle=(0.8\pm0.1)\langle\bar{q}q \rangle$,
 $\langle\bar{q}g_s\sigma G q \rangle=m_0^2\langle \bar{q}q \rangle$, $\langle\bar{s}g_s\sigma G s \rangle=m_0^2\langle \bar{s}s \rangle$,
$m_0^2=(0.8 \pm 0.1)\,\rm{GeV}^2$, $\langle \frac{\alpha_s
GG}{\pi}\rangle=0.012\pm0.004\,\rm{GeV}^4$    at the energy scale  $\mu=1\, \rm{GeV}$
\cite{SVZ79-1,SVZ79-2,PRT85,ColangeloReview}, and  take the $\overline{MS}$ quark  masses $m_{c}(m_c)=(1.275\pm0.025)\,\rm{GeV}$
 and $m_s(\mu=2\,\rm{GeV})=(0.095\pm0.005)\,\rm{GeV}$
 from the Particle Data Group \cite{PDG}.
In addition,  we take account of the energy-scale dependence of  those  input parameters from the re-normalization group equation with the lowest order approximation \cite{Narison-mix},
 \begin{eqnarray}
 \langle\bar{q}q \rangle(\mu)&=&\langle\bar{q}q\rangle({\rm 1 GeV})\left[\frac{\alpha_{s}({\rm 1 GeV})}{\alpha_{s}(\mu)}\right]^{\frac{12}{33-2n_f}}\, , \nonumber\\
 \langle\bar{s}s \rangle(\mu)&=&\langle\bar{s}s \rangle({\rm 1 GeV})\left[\frac{\alpha_{s}({\rm 1 GeV})}{\alpha_{s}(\mu)}\right]^{\frac{12}{33-2n_f}}\, , \nonumber\\
 \langle\bar{q}g_s \sigma Gq \rangle(\mu)&=&\langle\bar{q}g_s \sigma Gq \rangle({\rm 1 GeV})\left[\frac{\alpha_{s}({\rm 1 GeV})}{\alpha_{s}(\mu)}\right]^{\frac{2}{33-2n_f}}\, ,\nonumber\\
  \langle\bar{s}g_s \sigma Gs \rangle(\mu)&=&\langle\bar{s}g_s \sigma Gs \rangle({\rm 1 GeV})\left[\frac{\alpha_{s}({\rm 1 GeV})}{\alpha_{s}(\mu)}\right]^{\frac{2}{33-2n_f}}\, ,\nonumber\\
m_c(\mu)&=&m_c(m_c)\left[\frac{\alpha_{s}(\mu)}{\alpha_{s}(m_c)}\right]^{\frac{12}{33-2n_f}} \, ,\nonumber\\
m_s(\mu)&=&m_s({\rm 2GeV} )\left[\frac{\alpha_{s}(\mu)}{\alpha_{s}({\rm 2GeV})}\right]^{\frac{12}{33-2n_f}}\, ,\nonumber\\
\alpha_s(\mu)&=&\frac{1}{b_0t}\left[1-\frac{b_1}{b_0^2}\frac{\log t}{t} +\frac{b_1^2(\log^2{t}-\log{t}-1)+b_0b_2}{b_0^4t^2}\right]\, ,
\end{eqnarray}
  where $t=\log \frac{\mu^2}{\Lambda^2}$, $b_0=\frac{33-2n_f}{12\pi}$, $b_1=\frac{153-19n_f}{24\pi^2}$, $b_2=\frac{2857-\frac{5033}{9}n_f+\frac{325}{27}n_f^2}{128\pi^3}$,  $\Lambda_{QCD}=210\,\rm{MeV}$, $292\,\rm{MeV}$  and  $332\,\rm{MeV}$ for the flavors  $n_f=5$, $4$ and $3$, respectively  \cite{PDG}.

In this work, we study  the hidden-charm pentaquark states $udsc\bar{c}$ with the isospin $I=0$,  and choose the flavor numbers $n_f=4$, then evolve all those input parameters to a typical energy scale $\mu$, which satisfies  the modified energy scale formula,
\begin{eqnarray}
\mu &=&\sqrt{M_{P}^2-(2{\mathbb{M}}_c)^2}-{\mathbb{M}}_s \, ,
 \end{eqnarray}
 with the effective quark masses ${\mathbb{M}}_c$ and ${\mathbb{M}}_s$, which characterize the heavy degrees of freedom and light-flavor $SU(3)$ breaking effects, the  updated values are ${\mathbb{M}}_c=1.82\,\rm{GeV}$ and ${\mathbb{M}}_s=0.15\,\rm{GeV}$ respectively
 \cite{WangZG-Pcs4459-333,Pcs4338-mole-XWWang,WZG-tetraquark-Mc,
 Wang-tetra-formula,WangZG-mole-formula-1,WangZG-mole-formula-2,
WangZG-XQ-mole-EPJA,WangZG-IJMPA-2021,Wang-tetra-PRD-HC,Wang-tetra-NPB-HCss}.

In the QCD sum rules for  the  baryons  and  pentaquark states contain at least one valence heavy quark,  we usually choose the continuum threshold parameters as $\sqrt{s_0}=M_{gr}+ (0.5-0.8)\,\rm{GeV}$  \cite{WangZG-Pcs4459-333,Pcs4338-mole-XWWang,WangXW-Pc-mole-SCPMA,
WangXW-Pc-mole-IJMPA,Wang1508-EPJC,WangHuang-EPJC-1508-12,
 WangZG-EPJC-1509-12,WangZG-NPB-1512-32,WZG-penta-IJMPA,Wang-cc-baryon-penta},   where the subscript $gr$ represent the ground states.
 In Ref.\cite{WangZG-Pcs4459-333},  we choose the continuum threshold parameter   $\sqrt{s_0}= 5.15\pm0.10\,\rm{GeV}$, and examine  the possible assignment of the $P_{cs}(4459)$ as the $[ud][sc]\bar{c}$ ($0$, $0$, $0$, $\frac{1}{2}$)  state. Now we extend our previous works to study all the possible hidden-charm pentaquark states with zero isospin in the $J/\psi \Lambda$ mass spectrum.

 We obtain the  Borel  windows and continuum threshold parameters via tedious trial  and error, which are shown in Table \ref{Borel}. From the table, we can see clearly that the pole contributions are about $(40-60)\%$, the pole dominance criterion is satisfied and it is reliable  to extract the pentaquark masses, where
  the pole contributions are defined by,
\begin{eqnarray}
{\rm{pole}}&=&\frac{\int_{4m_{c}^{2}}^{s_{0}}ds\,\rho_{QCD}\left(s\right)\exp\left(-\frac{s}{T^{2}}\right)} {\int_{4m_{c}^{2}}^{\infty}ds\,\rho_{QCD}\left(s\right)\exp\left(-\frac{s}{T^{2}}\right)}\, ,
\end{eqnarray}
 with the spectral densities  $\rho_{QCD}=\sqrt{s}\rho_{QCD}^1(s)+\rho_{QCD}^{0}(s)$.

\begin{figure}
\centering
\includegraphics[totalheight=6cm,width=7cm]{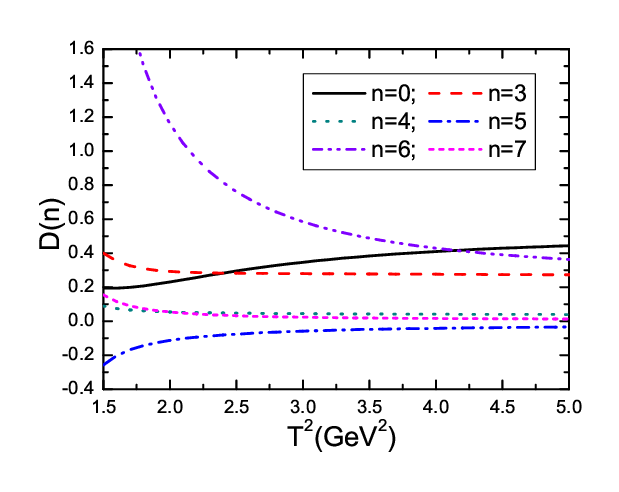}
\includegraphics[totalheight=6cm,width=7cm]{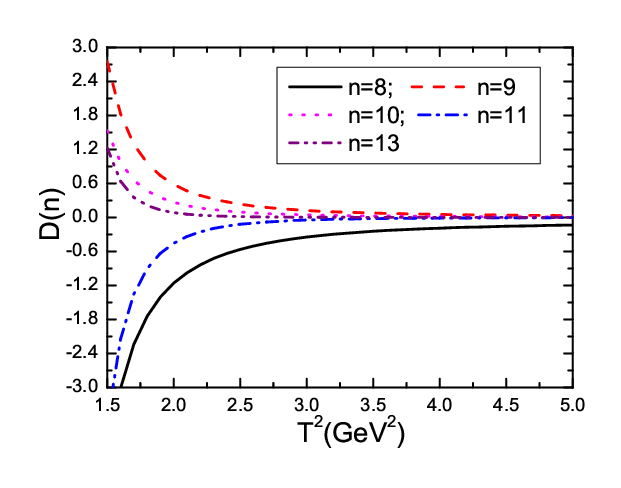}
\includegraphics[totalheight=6cm,width=7cm]{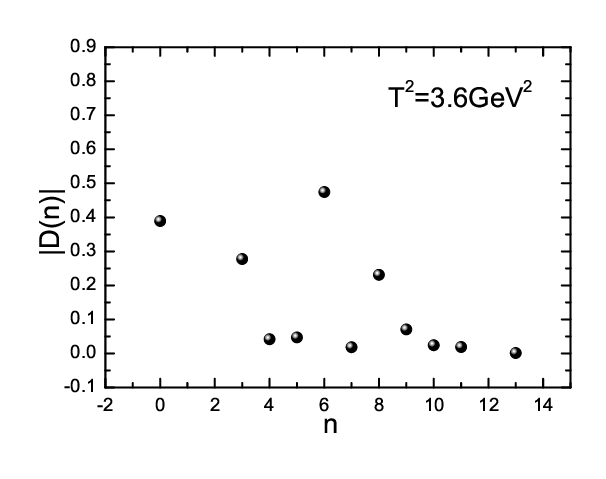}
  \caption{ The  contributions of the vacuum condensates $D(n)$ with variations of the  Borel parameter $T^2$ for the $[ud][sc]\bar{c}$ ($0$, $0$, $0$, $\frac{1}{2}$) pentaquark state. }\label{Fr-Pcs12-1}
\end{figure}

 In Fig.\ref{Fr-Pcs12-1}, we plot the contributions of the vacuum condensates of dimension $n$ ($D(n)$) with variations of the  Borel parameter $T^2$ for the $[ud][sc]\bar{c}$ ($0$, $0$, $0$, $\frac{1}{2}$) pentaquark state as an example, where the $D(n)$ are defined by,
   \begin{eqnarray}
D(n)&=&\frac{\int_{4m_{c}^{2}}^{s_{0}}ds\,\rho_{QCD,n}(s)\exp\left(-\frac{s}{T^{2}}\right)}
{\int_{4m_{c}^{2}}^{s_{0}}ds\,\rho_{QCD}\left(s\right)\exp\left(-\frac{s}{T^{2}}\right)}\, .
\end{eqnarray}
From the figure, we can see clearly that  in the whole region the $D(4)$, $D(5)$ and $D(7)$ play a tiny role, while the $D(6)$ plays an important role, it is unreliable to judge the convergent behavior of the operator product expansion by only considering the vacuum condensates up to dimension $7$.  At small value of the Borel parameter $T^2$, the $D(8)$, $D(9)$, $D(10)$, $D(11)$ and $D(13)$ manifest themselves significantly, thus they play an important role in determining the Borel windows. In fact, the $D(8)$ serves as a milestone, at the center   of the  Borel window $T^2=3.6\,\rm{GeV}^2$, see Fig.\ref{Fr-Pcs12-1}, the vacuum condensates have the hierarchy $|D(8)|\gg D(9)\gg D(10)\geq |D(11)| \geq D(13)$, the operator product expansion converges very well. Again, let us  look at Table \ref{Borel},  $D(13)\ll 1\%$ except for the current $J^3(x)$, where $D(13)<2\%$. All in all, the operator product expansion is convergent.

\begin{table}
\begin{center}
\begin{tabular}{|c|c|c|c|c|c|c|c|}\hline\hline
                  &$T^2(\rm{GeV}^2)$     &$\sqrt{s_0}(\rm{GeV})$    &$\mu(\rm{GeV})$  &pole          &$D(13)$         \\ \hline

$J^1(x)$          &$3.4-3.8$             &$5.15\pm0.10$             &$2.4$            &$(40-61)\%$   &$\ll 1\%$      \\ \hline

$J^2(x)$          &$3.4-3.8$             &$5.20\pm0.10$             &$2.5$            &$(41-62)\%$   &$\ll 1\%$       \\ \hline

$J^3(x)$          &$3.0-3.4$             &$5.00\pm0.10$             &$2.2$            &$(40-62)\%$   &$<2\%$      \\ \hline

$J^4(x)$          &$3.3-3.7$             &$5.05\pm0.10$             &$2.3$            &$(40-60)\%$   &$\ll1\%$     \\ \hline

$J^1_\mu(x)$      &$3.4-3.8$             &$5.20\pm0.10$             &$2.5$            &$(42-62)\%$   &$\ll 1\%$     \\ \hline

$J^2_\mu(x)$      &$3.4-3.8$             &$5.15\pm0.10$             &$2.4$            &$(41-61)\%$   &$\ll1\%$     \\ \hline

$J^3_\mu(x)$      &$3.4-3.8$             &$5.10\pm0.10$             &$2.4$            &$(40-60)\%$   &$\ll1\%$     \\ \hline

$J^4_\mu(x)$      &$3.4-3.8$             &$5.15\pm0.10$             &$2.4$            &$(40-60)\%$   &$\ll1\%$     \\ \hline

$J^5_\mu(x)$      &$3.4-3.8$             &$5.15\pm0.10$             &$2.4$            &$(40-60)\%$   &$\ll1\%$     \\ \hline

$J^1_{\mu\nu}(x)$ &$3.4-3.8$             &$5.20\pm0.10$             &$2.5$            &$(42-62)\%$   &$\ll1\%$     \\ \hline

$J^2_{\mu\nu}(x)$ &$3.5-3.9$             &$5.20\pm0.10$             &$2.5$            &$(40-60)\%$   &$\ll1\%$     \\ \hline\hline
\end{tabular}
\end{center}
\caption{ The Borel  windows, continuum threshold parameters, ideal energy scales, pole contributions,   contributions of the vacuum condensates of dimension 13 for the hidden-charm pentaquark states with zero isospin. }\label{Borel}
\end{table}

\begin{table}
\begin{center}
\begin{tabular}{|c|c|c|c|c|c|c|c|c|}\hline\hline
$[qq][qc]\bar{c}$ ($S_L$, $S_H$, $J_{LH}$, $J$) &$M(\rm{GeV})$   &$\lambda(10^{-3}\rm{GeV}^6)$ &Assignments       \\ \hline

$[ud][sc]\bar{c}$ ($0$, $0$, $0$, $\frac{1}{2}$)                      &$4.47\pm0.11$   &$1.86\pm0.30$                 &$?\,P_{cs}(4459)$         \\

$[ud][sc]\bar{c}$ ($0$, $1$, $1$, $\frac{1}{2}$)                      &$4.51\pm0.10$   &$3.43\pm0.55$                 &                       \\

$[us][dc]\bar{c}-[ds][uc]\bar{c}$ ($1$, $1$, $0$, $\frac{1}{2}$)     &$4.33\pm0.11$   &$2.34\pm0.42$                &$?\,P_{cs}(4338)$          \\

$[us][dc]\bar{c}-[ds][uc]\bar{c}$ ($1$, $0$, $0$, $\frac{1}{2}$)     &$4.37\pm0.11$   &$2.81\pm0.47$                &$??\,P_{cs}(4338)$        \\

$[ud][sc]\bar{c}$ ($0$, $1$, $1$, $\frac{3}{2}$)                      &$4.51\pm0.11$   &$1.87\pm0.30$                &                   \\

$[us][dc]\bar{c}-[ds][uc]\bar{c}$ ($0$, $1$, $1$, $\frac{3}{2}$)     &$4.46\pm0.10$   &$1.76\pm0.28$                &$?\,P_{cs}(4459)$     \\

$[us][dc]\bar{c}-[ds][uc]\bar{c}$ ($1$, $0$, $1$, $\frac{3}{2}$)     &$4.42\pm0.10$   &$1.68\pm0.27$                &$??\,P_{cs}(4459)$                     \\

$[us][dc]\bar{c}-[ds][uc]\bar{c}$ ($1$, $1$, $2$, $\frac{3}{2}$)${}_4$     &$4.47\pm0.10$   &$3.05\pm0.49$                &$?\,P_{cs}(4459)$    \\

$[us][dc]\bar{c}-[ds][uc]\bar{c}$ ($1$, $1$, $2$, $\frac{3}{2}$)${}_5$     &$4.47\pm0.10$   &$3.04\pm0.50$                &$?\,P_{cs}(4459)$     \\

$[ud][sc]\bar{c}$ ($0$, $1$, $1$, $\frac{5}{2}$)
&$4.51\pm0.10$   &$1.87\pm0.30$                &      \\

$[us][dc]\bar{c}-[ds][uc]\bar{c}$ ($1$, $1$, $2$, $\frac{5}{2}$)                      &$4.51\pm0.10$   &$1.81\pm0.28$                &     \\ \hline\hline
\end{tabular}
\end{center}
\caption{ The masses  and pole residues of the hidden-charm pentaquark states with possible assignments. }\label{mass-Pcs}
\end{table}

\begin{figure}
\centering
\includegraphics[totalheight=6cm,width=7cm]{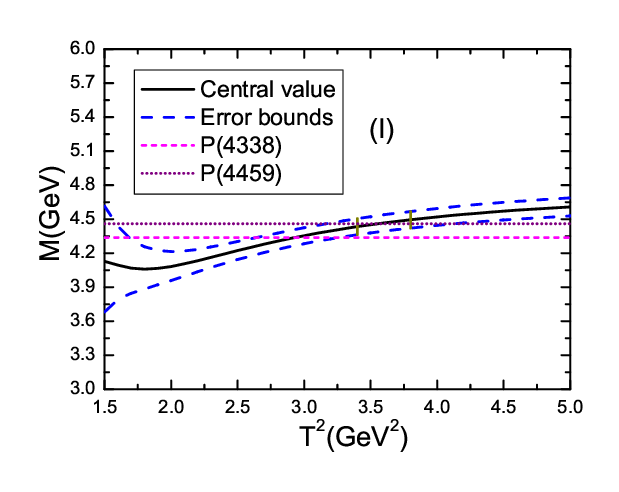}
\includegraphics[totalheight=6cm,width=7cm]{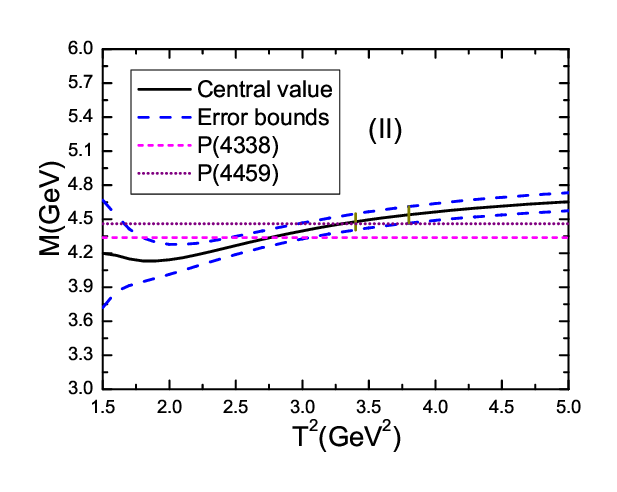}
\includegraphics[totalheight=6cm,width=7cm]{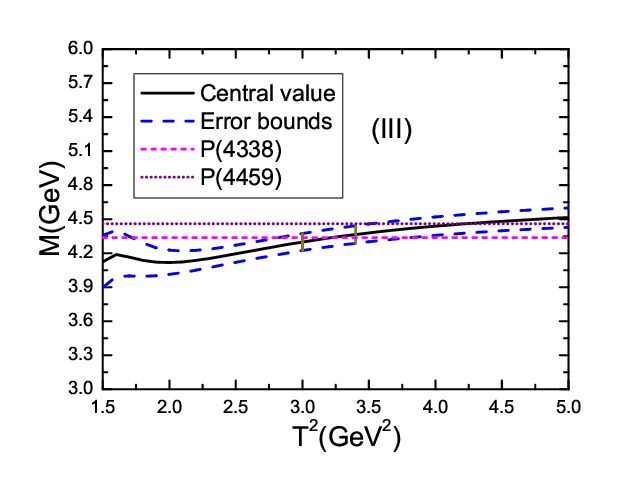}
\includegraphics[totalheight=6cm,width=7cm]{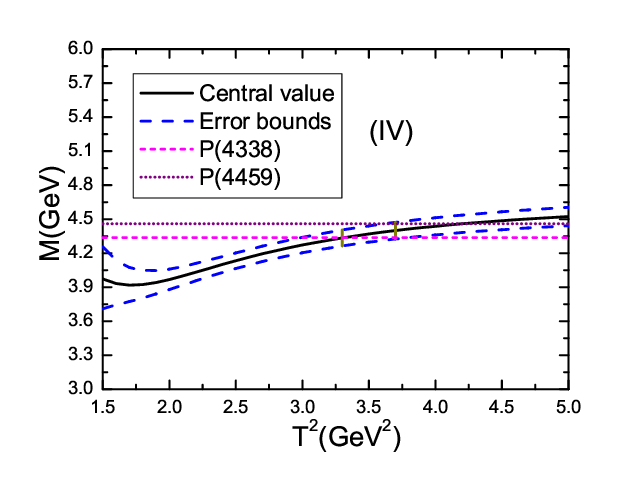}
  \caption{ The masses  with variations of the  Borel parameters $T^2$ for  the hidden-charm pentaquark states, where the (I), (II), (III)  and (IV)  denote the
   $[ud][sc]\bar{c}$ ($0$, $0$, $0$, $\frac{1}{2}$),                    $[ud][sc]\bar{c}$ ($0$, $1$, $1$, $\frac{1}{2}$),
$[us][dc]\bar{c}-[ds][uc]\bar{c}$ ($1$, $1$, $0$, $\frac{1}{2}$)   and
$[us][dc]\bar{c}-[ds][uc]\bar{c}$ ($1$, $0$, $0$, $\frac{1}{2}$)  pentaquark states, respectively. }\label{mass-1-fig}
\end{figure}

\begin{figure}
\centering
\includegraphics[totalheight=6cm,width=7cm]{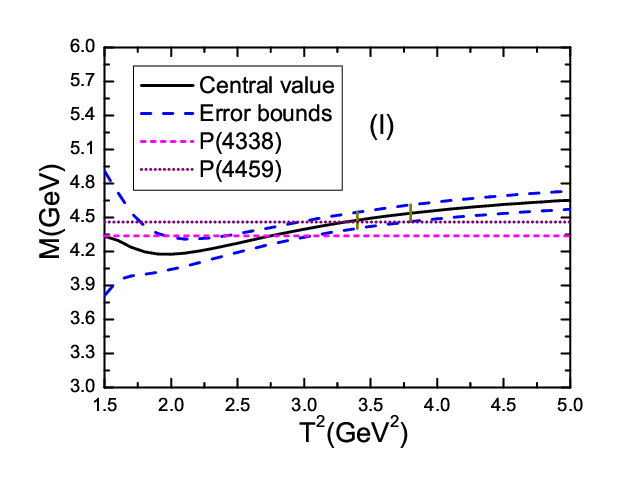}
\includegraphics[totalheight=6cm,width=7cm]{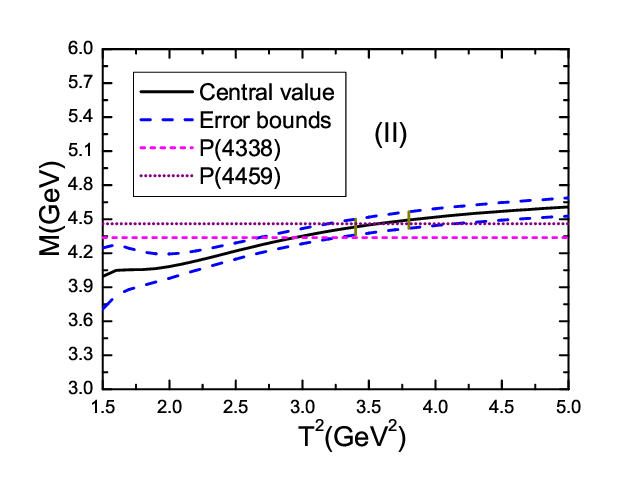}
\includegraphics[totalheight=6cm,width=7cm]{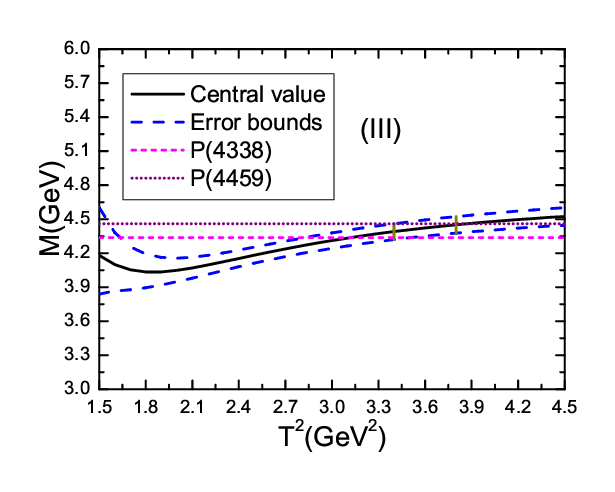}
\includegraphics[totalheight=6cm,width=7cm]{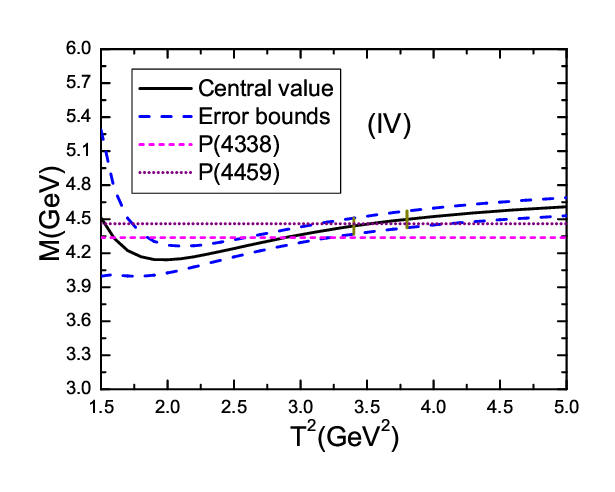}
\includegraphics[totalheight=6cm,width=7cm]{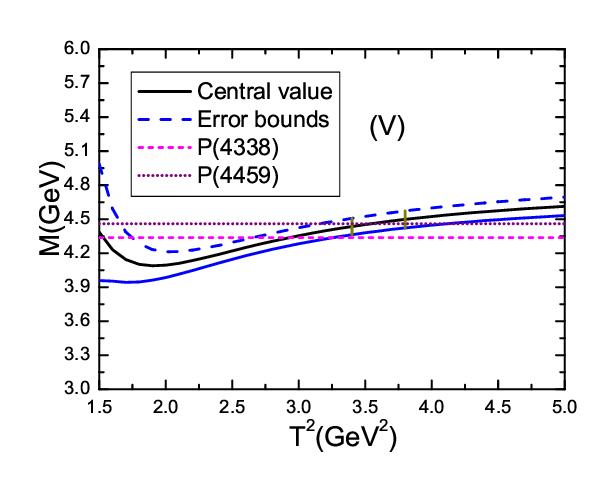}
  \caption{ The masses  with variations of the  Borel parameters $T^2$ for  the hidden-charm pentaquark states, where the (I), (II), (III), (IV) and (V) denote the $[ud][sc]\bar{c}$ ($0$, $1$, $1$, $\frac{3}{2}$),
$[us][dc]\bar{c}-[ds][uc]\bar{c}$ ($0$, $1$, $1$, $\frac{3}{2}$),
$[us][dc]\bar{c}-[ds][uc]\bar{c}$ ($1$, $0$, $1$, $\frac{3}{2}$),
$[us][dc]\bar{c}-[ds][uc]\bar{c}$ ($1$, $1$, $2$, $\frac{3}{2}$)${}_4$   and
$[us][dc]\bar{c}-[ds][uc]\bar{c}$ ($1$, $1$, $2$, $\frac{3}{2}$)${}_5$ pentaquark states, respectively.  }\label{mass-2-fig}
\end{figure}

\begin{figure}
\centering
\includegraphics[totalheight=6cm,width=7cm]{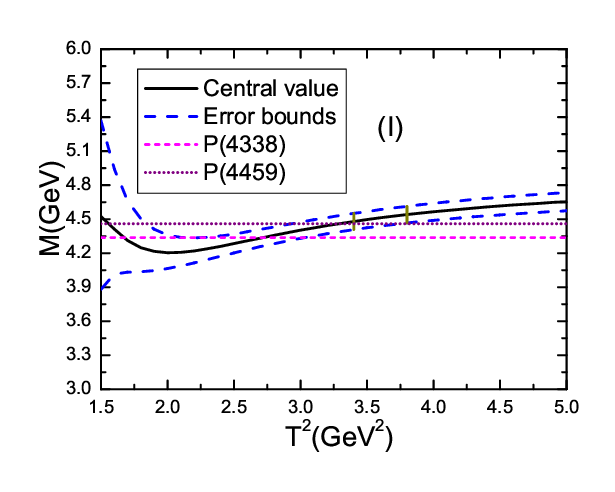}
\includegraphics[totalheight=6cm,width=7cm]{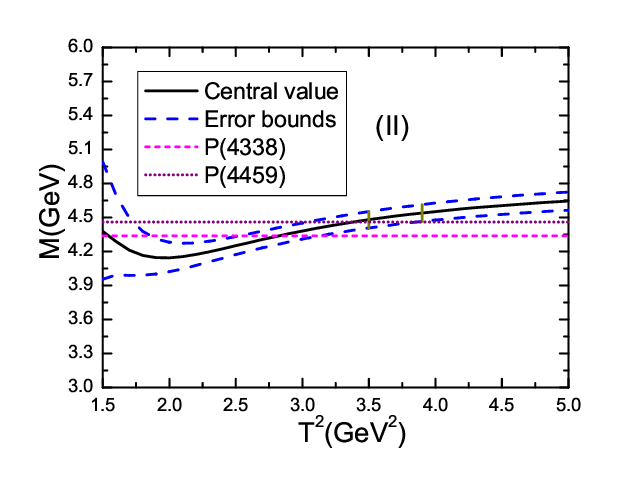}
  \caption{ The masses  with variations of the  Borel parameters $T^2$ for  the hidden-charm pentaquark states, where the (I) and (II)  denote the
  $[ud][sc]\bar{c}$ ($0$, $1$, $1$, $\frac{5}{2}$)
and $[us][dc]\bar{c}-[ds][uc]\bar{c}$ ($1$, $1$, $2$, $\frac{5}{2}$)
    pentaquark states, respectively. }\label{mass-3-fig}
\end{figure}

\begin{figure}
\centering
\includegraphics[totalheight=6cm,width=7cm]{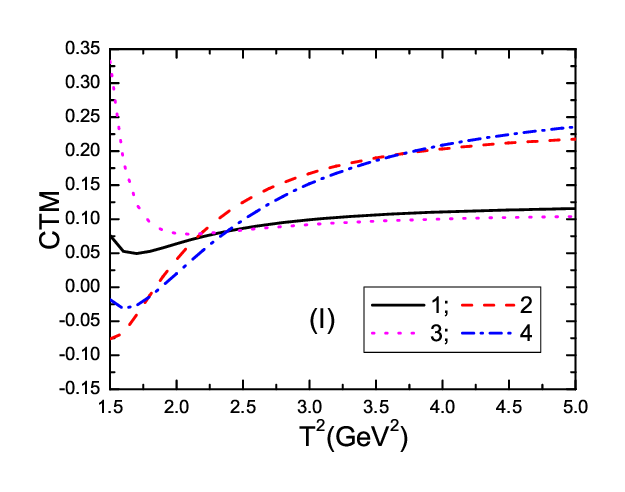}
\includegraphics[totalheight=6cm,width=7cm]{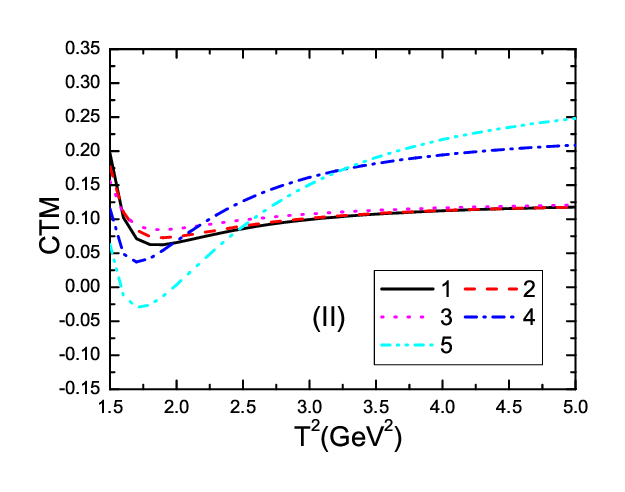}
\includegraphics[totalheight=6cm,width=7cm]{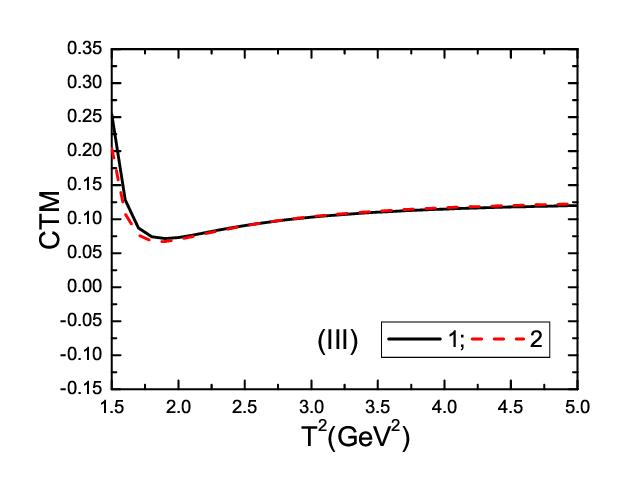}
  \caption{ The parameters CTM measuring  contributions from the hidden-charm pentaquark states with the positive parity, where the (I), (II) and (III) denote  the spin $J=\frac{1}{2}$, $\frac{3}{2}$ and $\frac{5}{2}$, the $1$, $2$, $3$, $4$ and $5$ denote the series numbers of the currents. }\label{CTM-Borel}
\end{figure}

Now we take  account of  all uncertainties  of the input   parameters,
and obtain  the masses and pole residues of
 the   hidden-charm pentaquark states, which are shown explicitly in Figs.\ref{mass-1-fig}-\ref{mass-3-fig} and Table \ref{mass-Pcs}. From Tables \ref{Borel}-\ref{mass-Pcs}, we can see that the modified energy scale formula
 $\mu =\sqrt{M^2_{P}-(2{\mathbb{M}}_c)^2}-{\mathbb{M}}_s$ is satisfied very well.
 The formula can enhance the pole contributions significantly and improve the convergent behavior of the operator product expansion  significantly \cite{WangZG-Review,WangZG-IJMPA-3-scheme}. Without adopting the energy scale formula, we could only obtain poor pole contributions and bad convergent behavior of the operator product expansion \cite{WangZG-IJMPA-3-scheme}.

In Figs.\ref{mass-1-fig}-\ref{mass-3-fig}, we plot the masses of the hidden-charm pentaquark states with   zero isospin, where the regions between the two vertical lines are the Borel windows. In the Borel windows, there appear flat platforms indeed. In those figures, we also present the experimental values of the masses of the $P_{cs}(4459)$ and $P_{cs}(4338)$ from the LHCb collaboration \cite{LHCb-Pcs4459-2012,LHCb-Pcs4338}, thus we could  obtain intuitive conclusions  about the possible assignments of the two $P_{cs}$ states.

The predicted mass  $M_P=4.33\pm0.11\,\rm{GeV}$ for the
$[us][dc]\bar{c}-[ds][uc]\bar{c}$ ($1$, $1$, $0$, $\frac{1}{2}$) pentaquark state
is in excellent agreement with the experimental data $4338.2\pm0.7\pm0.4\,\rm{MeV}$ from the  LHCb collaboration \cite{LHCb-Pcs4338}, and supports assigning the
    $P_{cs}(4338)$ as the $[us][dc]\bar{c}-[ds][uc]\bar{c}$ ($1$, $1$, $0$, $\frac{1}{2}$) pentaquark state with the spin-parity $J^P={\frac{1}{2}}^-$, the favored spin-parity of the $P_{cs}(4338)$.
While the predicted mass $M_P=4.37\pm0.11\,\rm{GeV}$ for the
$[us][dc]\bar{c}-[ds][uc]\bar{c}$ ($1$, $0$, $0$, $\frac{1}{2}$) pentaquark state is somewhat larger than  the experimental data $4338.2\pm0.7\pm0.4\,\rm{MeV}$ from the  LHCb collaboration \cite{LHCb-Pcs4338}, it is marginal to assign the $P_{cs}(4338)$ as the  $[us][dc]\bar{c}-[ds][uc]\bar{c}$ ($1$, $0$, $0$, $\frac{1}{2}$) pentaquark state  with the spin-parity $J^P={\frac{1}{2}}^-$.

The predicted masses  $M_P=4.47\pm0.11\,\rm{GeV}$, $4.46\pm0.10\,\rm{GeV}$,  $4.47\pm0.10\,\rm{GeV}$ and $4.47\pm0.10\, \rm{GeV}$
for the
$[ud][sc]\bar{c}$ ($0$, $0$, $0$, $\frac{1}{2}$),
$[us][dc]\bar{c}-[ds][uc]\bar{c}$ ($0$, $1$, $1$, $\frac{3}{2}$),
$[us][dc]\bar{c}-[ds][uc]\bar{c}$ ($1$, $1$, $2$, $\frac{3}{2}$)${}_4$ and
$[us][dc]\bar{c}-[ds][uc]\bar{c}$ ($1$, $1$, $2$, $\frac{3}{2}$)${}_5$     pentaquark states  are all in excellent agreement with the experimental data $ 4458.8 \pm 2.9 {}^{+4.7}_{-1.1} \mbox{ MeV}$ from the  LHCb collaboration  \cite{LHCb-Pcs4459-2012}, and supports assigning the $P_{cs}(4459)$   as the hidden-charm pentaquark state
with the spin-parity $J^P={\frac{1}{2}}^-$ or ${\frac{3}{2}}^-$.
While the predicted mass $M_P=4.42\pm0.10\,\rm{GeV}$ for the
$[us][dc]\bar{c}-[ds][uc]\bar{c}$ ($1$, $0$, $1$, $\frac{3}{2}$) pentaquark state is somewhat lower than the experimental data $ 4458.8 \pm 2.9 {}^{+4.7}_{-1.1} \mbox{ MeV}$ from the  LHCb collaboration  \cite{LHCb-Pcs4459-2012}, it is marginal to assigning the $P_{cs}(4459)$ as the  $[us][dc]\bar{c}-[ds][uc]\bar{c}$ ($1$, $0$, $1$, $\frac{3}{2}$) pentaquark state with the spin-parity $J^P={\frac{3}{2}}^-$. All in all, there are enough rooms to accommodate the  two $P_{cs}$ states in the scenario of  pentaquark states.
As we cannot assign a hadron based on the mass alone unambiguously, we should study its production, decays, etc in a comprehensive way. We can take the pole residues as basic input parameters and study the two-body strong decays,
\begin{eqnarray}
P_{cs}&\to& \bar{D}\Xi_c\, , \, \bar{D}_s\Lambda_c\, , \, \bar{D}^*\Xi_c\, , \, \bar{D}_s^*\Lambda_c\, , \, J/\psi \Lambda \, , \, \eta_c \Lambda \, ,
\end{eqnarray}
 with the three-point QCD sum rules to estimate the decay widths and select the optimal channels to search for those pentaquark states.
Recently, the LHCb collaboration observed the
$\Lambda_b^{0} \to \Lambda_{c}^{+}D_{s}^{-}K^{+}K^{-}$ decay for the first time  and found no evidence of the  pentaquark candidates $P_{cs}(4338)$ and $P_{cs}(4459)$ in the $\Lambda_{c}^{+}D_{s}^{-}$ mass spectrum \cite{LHCb-Lambda-Ds}.

In Fig.\ref{CTM-Borel}, we plot the parameters CTM measuring  contributions from the hidden-charm pentaquark states with positive parity with variations of the Borel parameters. From the figure, we can see that ${\rm CTM}\sim 0.10$ or $0.20$ in the Borel windows, the contaminations from the hidden-charm pentaquark states with positive parity are considerable if the two traditional  QCD sum rules in Eqs.\eqref{Traditional-QCDSR-1}-\eqref{Traditional-QCDSR-0} are adopted.

\section{Conclusion}
 In this work, we distinguish the isospin  for the first time and select the isospin zero configurations to  study the diquark-diquark-antiquark type $udsc\bar{c}$ pentaquark states in the framework of  the QCD sum rules systematically. We take account of   the    vacuum condensates up to dimension $13$ consistently, obtain the QCD spectral densities and distinguish the contributions from the pentaquark states with the negative and positive parity unambiguously,  then adopt  the modified energy scale formula $\mu=\sqrt{M_{P}-(2{\mathbb{M}}_c)^2}-{\mathbb{M}}_s$ to choose  the optimal  energy scales of the QCD spectral densities to enhance the pole contributions and improve the convergent behavior of the operator product expansion. Finally, we obtain the mass spectrum of the $udsc\bar{c}$ pentaquark states with the quantum numbers $I=0$ and $J^{P}={\frac{1}{2}}^-$, ${\frac{3}{2}}^-$, ${\frac{5}{2}}^-$.
 The present predictions support assigning the
    $P_{cs}(4338)$ as the $[us][dc]\bar{c}-[ds][uc]\bar{c}$ ($1$, $1$, $0$, $\frac{1}{2}$) pentaquark state with the spin-parity $J^P={\frac{1}{2}}^-$,
 assigning the $P_{cs}(4459)$ as the $[ud][sc]\bar{c}$ ($0$, $0$, $0$, $\frac{1}{2}$) pentaquark state with the spin-parity $J^P={\frac{1}{2}}^-$, or
$[us][dc]\bar{c}-[ds][uc]\bar{c}$ ($0$, $1$, $1$, $\frac{3}{2}$),
$[us][dc]\bar{c}-[ds][uc]\bar{c}$ ($1$, $1$, $2$, $\frac{3}{2}$)${}_4$,
$[us][dc]\bar{c}-[ds][uc]\bar{c}$ ($1$, $1$, $2$, $\frac{3}{2}$)${}_5$ pentaquark state with the spin-parity $J^P={\frac{3}{2}}^-$. More experimental data are still needed to make an unambiguous assignment.

\section*{Acknowledgements}
This  work is supported by National Natural Science Foundation, Grant Number  12575083.

\end{document}